\documentclass[a4paper,aps,prd,10pt,preprintnumbers,showpacs,twocolumn,superscriptaddress,nofootinbib,amsmath,amssymb]{revtex4-1}
\usepackage[dvips]{graphics}
\usepackage[utf8]{inputenc}
\usepackage[T1]{fontenc}
\usepackage{cmap}
\def\imo{i}
\def\re#1{Re(#1)}
\def\im#1{Im(#1)}
\def\K{{\cal K}}

\begin{document}
\title[Higher order WKB formula for quasinormal modes and grey-body factors]{Higher order WKB formula for quasinormal modes and grey-body factors: recipes for quick and accurate calculations}
\author{R. A. Konoplya}\email{konoplya\_roma@yahoo.com}
\affiliation{Institute of Physics and Research Centre of Theoretical Physics and Astrophysics, Faculty of Philosophy and Science, Silesian University in Opava, CZ-746 01 Opava, Czech Republic}
\affiliation{Peoples Friendship University of Russia (RUDN University), 6 Miklukho-Maklaya Street, Moscow 117198, Russian Federation}
\author{A. Zhidenko}\email{olexandr.zhydenko@ufabc.edu.br}
\affiliation{Institute of Physics and Research Centre of Theoretical Physics and Astrophysics, Faculty of Philosophy and Science, Silesian University in Opava, CZ-746 01 Opava, Czech Republic}
\affiliation{Centro de Matemática, Computação e Cognição (CMCC), Universidade Federal do ABC (UFABC),\\ Rua Abolição, CEP: 09210-180, Santo André, SP, Brazil}
\author{A. F. Zinhailo}\email{F170631@fpf.slu.cz}
\affiliation{Institute of Physics and Research Centre of Theoretical Physics and Astrophysics, Faculty of Philosophy and Science, Silesian University in Opava, CZ-746 01 Opava, Czech Republic}
\begin{abstract}
The WKB approach for finding quasinormal modes of black holes, suggested in \cite{1985ApJ...291L..33S} by Schutz and Will at the first order and later developed to higher orders \cite{Iyer:1986np,Konoplya:2003ii,Matyjasek:2017psv}, became popular during the past decades, because, unlike more sophisticated numerical approaches, it is automatic for different effective potentials and mostly provides sufficient accuracy. At the same time, the seeming simplicity of the WKB approach resulted in appearance of a big number of partially misleading papers, where the WKB formula was used beyond its scope of applicability. Here we review various situations in which the WKB formula can or cannot bring us to reliable conclusions. As the WKB series converges only asymptotically, there is no mathematically strict criterium for evaluation of an error. Therefore, here we are trying to introduce a number of practical recipes instead and summarize cases in which higher WKB orders improve accuracy. We show that averaging of the Padé approximations, suggested first by J.~Matyjasek and M.~Opala \cite{Matyjasek:2017psv}, leads to much higher accuracy of the WKB approach, estimate the error and present the automatic code \cite{Mathematica-code} which computes quasinormal modes and grey-body factors.
\end{abstract}
\maketitle

\section{Introduction}
Quasinormal modes are characteristics of the late time response of a black hole to perturbation. Dominant quasinormal modes are clearly seen in the gravitational wave signal from black holes or other compact objects at late times. Therefore, they have been observed in recent series of experiments by LIGO/VIRGO collaborations \cite{TheLIGOScientific:2016src}. Calculations of quasinormal modes with high accuracy is an important task aimed at constraining possible gravitational theories and testing the regime of strong gravity.

There are a few numerical approaches designed to find quasinormal modes with any desired accuracy (see~\cite{Konoplya:2011qq} for a review of methods), which are based on convergent procedures. However, they require analysis of singular points of master differential equations, representing wave dynamics of a compact object. This analysis is frequently non-trivial and, what is more important, different for different spacetimes. Therefore, each time a numerical procedure must be developed separately, case by case.  Therefore, an automatic procedure which on one side would be unaltered for various master wave equations and on the other provide sufficient accuracy were appealing.

\begin{table}
\begin{tabular}{|c|c|}
   \hline
  WKB order & Publication \\
   \hline
  1 &  B. Mashhoon \cite{Mashhoon}, B. Schutz and C. Will \cite{1985ApJ...291L..33S}\\
  \hline
  2-3 & S. Iyer and C. Will \cite{Iyer:1986np} \\
   \hline
  4-6 & R. Konoplya \cite{Konoplya:2003ii} \\
   \hline
  7-13 & J.~Matyjasek and M.~Opala \cite{Matyjasek:2017psv} \\
  \hline
    \hline
\end{tabular}
\caption{WKB formula for the potential barrier problem with two turning points at different orders.}\label{tabl:WKBorders}
\end{table}
The first simple semi-analytic formula for finding quasinormal modes was suggested by B.~Mashhoon~\cite{Mashhoon} in~1983. It was based on the matching the effective potential by the inverse Pöschl-Teller potential~\cite{Poschl-Teller}, but the accuracy for lower multipole numbers was rather bad. A fruitful idea was developed by B.~Schutz and~C.~Will a year later. They used the WKB expansions at both asymptotic regions, near the event horizon and at infinity, and matched them near the peak of the effective potential with the Taylor expansion. At the first WKB order the Schutz-Will formula reproduced the Mashhoon formula. Surprisingly, already at the first WKB order~\cite{1985ApJ...291L..33S} it estimated the dominant gravitational quasinormal modes with the accuracy of about $6\%$, while the third order extension by S.~Iyer and~C.~Will~\cite{Iyer:1986np} improved the accuracy for the fundamental mode up to fractions of one percent. The WKB formula developed in  \cite{1985ApJ...291L..33S} and \cite{Iyer:1986np}  works well once $\ell\gg n$, where $\ell$ and $n$ are the multipole and overtone numbers respectively. Therefore, the fact that the lowest dynamical mode for gravitational perturbation of the Schwarzschild black hole corresponds to $\ell=2$ was a lucky moment, which was lost once more complex configurations, for example with a scalar field, were taken into consideration \cite{Konoplya:2001ji}. Thus, for $\ell=0$, $n=0$ modes corresponding to perturbations of a scalar field the relative error was about ten percents at the third WKB order. Extension of the formula to the sixth order allowed to diminish the relative error for a number of cases by quite a few times or even orders \cite{Konoplya:2003ii}. Nevertheless, in many cases the WKB formula did not allow one to compute $n\geq\ell$ modes with satisfactory accuracy and it was not clear whether further extension in orders will be effective, as the WKB series is known to converge only asymptotically. Therefore, alternative semianalytical approaches, such as phase-integral treatment \cite{Froeman:1992gp,Andersson:1992scr}, were proposed.

Despite WKB approach frequently works better than it is expected, it does not guarantee convergence in each order. The great increase of the accuracy of WKB approach has been recently provided by the usage of the Padé approximation (by J.~Matyjasek and M.~Opala~\cite{Matyjasek:2017psv}) which helps to guess the asymptotic behavior of the WKB series. In \cite{Galtsov:1991nwq} the WKB approach was extended to the case of three turning points, which included a massive scalar field, while in \cite{Simone:1991wn} the WKB formula was applied to a massive scalar field in the Schwarzschild and Kerr backgrounds for the first time. The reflection/transmission coefficients for the scattering problem in various backgrounds were found with the help of the higher order WKB method in \cite{Konoplya:2010kv}. The summary of publications on developing the orders of the WKB formula is given in table~\ref{tabl:WKBorders}.

The WKB formula developed and extended in the above works have been applied to finding of quasinormal modes of black holes and other compact objects in hundreds of publications. Only the list of the past few years includes about a hundred works (see, for example \cite{Fernando:2016ftj,Konoplya:2016pmh,Molina:2016tkr,Cuyubamba:2016cug,Abbasvandi:2016yos,Toshmatov:2016bsb,Chen:2016qii,Konoplya:2016hmd,Fernando:2017qrd,Wahlang:2017zvk,Breton:2017hwe,Toshmatov:2017bpx,Toshmatov:2017qrq,Ciric:2017rnf,Burikham:2017gdm,Ovgun:2017dvs,Blazquez-Salcedo:2017bld,Chen:2017kqa,Panotopoulos:2017hns,MoraisGraca:2017hrf,Ponglertsakul:2018smo,Wu:2018xza,Toshmatov:2018tyo,Konoplya:2018ala,Panotopoulos:2018can,Blazquez-Salcedo:2018ipc,Chen:2018mkf,Panotopoulos:2018hua,Das:2018fzc,Dey:2018cws,Zinhailo:2018ska,Ge:2018vjq,Chakrabarty:2018skk,Chakrabarti:2018aqm,Oliveira:2018oha,Ding:2019tvs,Chen:2019iuo,Volkel:2019ahb,Konoplya:2019hml,Ciric:2019uab,Konoplya:2019ppy,Pramanik:2019qgy} and references therein). At the same time among these publications there are a lot of works in which the WKB method is either not properly used or misleading conclusions made from the performed computations. One of the most popular mistakes is when an observed effect is smaller or of the same order as an expected error of the WKB formula. Another common mistake is to claim that the obtained WKB spectrum of an object proves the stability. Sometimes the WKB formula developed for effective potentials with two turning points is applied to those with three and more turning points, what sometimes lead to much less accurate or even misleading results.

Therefore, in the present paper we would like to consider the state-of-art of the WKB approach suggested by B.~Schutz and~C.~Will, review plenty of qualitatively different situations, such as superradiance, quasiresonances, instability, charged fields and higher dimensions and overtones, etc., in which WKB formula can or cannot be used. Also we give a number of practical recipes for usage of the WKB approach, error estimation, and present the automatic \emph{Mathematica®} code \cite{Mathematica-code} which finds quasinormal modes and grey-body factors at a required WKB order up to 13th.

The paper is organized as follows. In Sec.~\ref{sec:WKB} we outline general properties of the WKB formula for quasinormal modes and transmission/reflection coefficients. Sec.~\ref{sec:spherical} is devoted to application of the above formula to static spherically symmetric black holes. Sec.~\ref{sec:Pade} discusses further improvement of the WKB approach by using Padé approximants. In Sec.~\ref{sec:using} we review situations in which WKB formula cannot be applied or should be used with reservations. These are: analysis of instability, calculations of the long-lived quasinormal modes of massive fields (quasiresonances), superradiance, calculations of higher overtones, probing spacetimes with non-constant asymptotics of the effective potentials. In Sec.~\ref{sec:massivescalar} we apply the WKB formula at higher orders to a number of cases studied before and review our observations related to the accuracy of the method. We also discuss in details possible ways of the error estimation. Finally, in Sec.~\ref{sec:summary} we summarize the obtained results and mention important open questions.

\section{WKB formula}\label{sec:WKB}

\begin{figure}
\resizebox{\linewidth}{!}{\includegraphics*{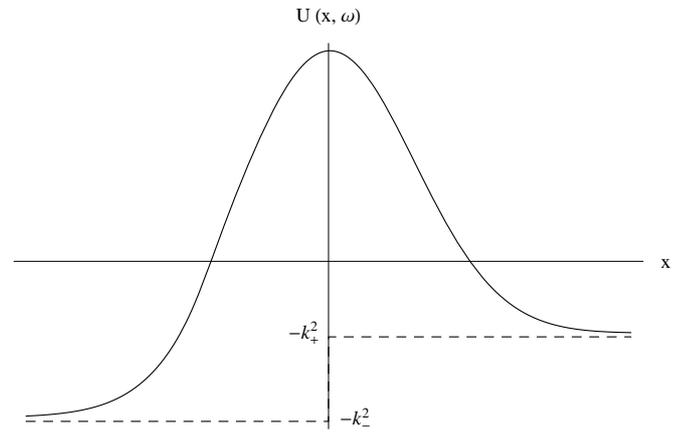}}
\caption{Effective potential with a single barrier and negative asymptotics.}\label{fig:pot}
\end{figure}

The WKB formula is appropriate for solving a wavelike equation,
\begin{equation}\label{wavelike}
  \frac{d^2\Psi}{dx^2}=U(x,\omega)\Psi,
\end{equation}
with the effective potential $U(x,\omega)$, which depends on a frequency of the wave $\omega\neq0$\footnote{We do not consider static solutions here.} and has a form of potential barrier with a single peak, approaching negative constants as $x\to\pm\infty$ (see fig.~\ref{fig:pot}). When the effective potential is asymptotically constant, any solution to the equation (\ref{wavelike}) in the asymptotic regions is a superposition of the ingoing and outgoing waves.

Conventionally, we shall assume that the time dependence of the perturbation function is given by the factor $\propto e^{-\imo\omega t}$. Hence, the wave is ingoing, when
\begin{equation}\label{ingoingwave}
\Psi_{in}(x\to\pm\infty)\propto \left\{
\begin{array}{ll}
 e^{-\imo k_{\pm} x}, & \omega>0; \\
 e^{\imo k_{\pm} x}, & \omega<0;
\end{array}
\right.
\end{equation}
and the wave is outgoing, when
\begin{equation}\label{outgoingwave}
\Psi_{out}(x\to\pm\infty)\propto \left\{
\begin{array}{ll}
 e^{\imo k_{\pm}(\omega) x}, & \omega>0; \\
 e^{-\imo k_{\pm}(\omega) x}, & \omega<0;
\end{array}
\right.
\end{equation}
where the asymptotic wave numbers $k_\pm(\omega)$ are positive, satisfying the dispersion relations
$$k_\pm^2(\omega)=-\lim_{x\to\pm\infty}U(x,\omega).$$

When studying black-hole perturbations the wavelike equation (\ref{wavelike}) is usually obtained with respect to the so-called tortoise coordinate, which is defined in such a way that $x\to\infty$ corresponds to spatial infinity and $x\to-\infty$ to the event horizon.

The WKB formula is based on matching of the asymptotic solutions, each of them being a superposition of (\ref{ingoingwave}) and (\ref{outgoingwave}), with the Taylor expansion around the top of the potential barrier $x=x_{0}$ through the two turning points, which are zeroes of the effective potential $U(x,\omega)$. In this way, it is possible to relate the ingoing and outgoing amplitudes through the linear transformation, i.~e. calculate $S$-matrix, which depends on the value of the potential in its maximum $U_0(\omega)= U(\omega, x_0)$,  and its higher derivatives\footnote{The first derivative $U_1(\omega)=0$ in the maximum.},
$$U_2(\omega) = \frac{d^2 U(\omega, x)}{d x^2}\Biggr|_{x=x_0},U_3(\omega) = \frac{d^3 U(\omega, x)}{d x^3}\Biggr|_{x=x_0},\ldots$$ (see e.~g. \cite{Konoplya:2011qq} for details). Finally, one finds that
\begin{eqnarray}\label{WKBformula}
&&0=U_0(\omega)+A_2(\K^2)+A_4(\K^2)+A_6(\K^2)+\ldots\\\nonumber&&-\imo \K\sqrt{-2U_2(\omega)}\left(1+A_3(\K^2)+A_5(\K^2)+A_7(\K^2)\ldots\right),
\end{eqnarray}
where $A_k(\K^2)$ is the correction of order $k$ to the eikonal formula.
Using (\ref{WKBformula}) one can calculate the reflection and transmission coefficients, $R$ and $T$, for the scattering problem defined as follows:
\begin{equation}\label{BC}
\begin{array}{ccll}
    \Psi &=& \Psi_{in}(x) + R \Psi_{out}(x),& x \rightarrow +\infty, \\
    \Psi &=& T \Psi_{in}(x),& x \rightarrow -\infty. \\
\end{array}%
\end{equation}

In particular, when the effective potential is real, $\K$ is a purely imaginary constant related with the reflection and transmission coefficients in the following way \cite{Iyer:1986np},
\begin{eqnarray}\label{reflection}
  |R|^2&=&\frac{1}{1+e^{-2\pi\imo \K}},\qquad 0<|R|^2<1.\\
\label{transmission}
  |T|^2&=&\frac{1}{1+e^{2\pi\imo \K}}=1-|R|^2.
\end{eqnarray}

The corrections $A_k(\K^2)$ are polynomials of the derivatives $U_2,U_3,\ldots U_{2k}$, and $\K^2$ with rational coefficients, divided by the appropriate power of $U_2$ and do not depend on $U_0$.\footnote{The denominator of $A_k$ is proportional to $(U_2)^m$, where $m$ is a positive integer, $m=\left[\frac{5(k-1)}{2}\right].$} Their explicit form can be found in \cite{Mathematica-code}.

It is important to note here that the eikonal formula provides the unique solution for $\K$,
\begin{equation}\label{eikonalK}
\K=-\imo \frac{U_0(\omega)}{\sqrt{-2U_2(\omega)}},
\end{equation}
so that the higher-order terms should be considered as corrections to this eikonal formula. We notice that equation~(\ref{WKBformula}) of order $k$ has precisely $k$ solutions for a given $\omega$, which determine $\K$.  The solution which represents the correction to the eikonal formula is usually given by the nearest root to that of the eikonal formula.

Quasinormal modes are eigenfrequencies $\omega$, for which the solution is a purely outgoing wave at spatial infinity and purely ingoing wave one the event horizon \cite{Konoplya:2011qq}. Quasinormal modes can be obtained within the WKB approach by analytic continuation of the $S$-matrix to the complex plane. The complex eigenfrequencies correspond to the poles of $\Gamma\left(-\K+\frac{1}{2}\right)$ for $\re{\omega}>0$ and $\Gamma\left(\K+\frac{1}{2}\right)$ for $\re{\omega}<0$ \cite{Iyer:1986np}. Thus, quasinormal modes can be found from (\ref{WKBformula}) by substituting half integer values of $\K$
\begin{eqnarray}\label{QNMsK}
\K &=& \left\{
\begin{array}{ll}
 +n+\frac{1}{2}, & \re{\omega}>0; \\
 -n-\frac{1}{2}, & \re{\omega}<0; \phantom{\frac{{}^{Whitespace}}{}}
\end{array}
\right.\\\nonumber
&&\qquad\quad\qquad n=0,1,2,3\ldots.
\end{eqnarray}
Note that, when $\K$ is given by (\ref{QNMsK}), denominators of (\ref{reflection}) and (\ref{transmission}) vanish, which is a consequence of the fact that the quasinormal modes are poles of the reflection and transmission coefficients.

In the general case one can compute the quasinormal frequencies using the formula~(\ref{WKBformula}). First, one should fix all the parameters in the effective potential. Then
one can find the value of $x_0$ at which $U$ attains maximum as a numerical function of $\omega$ and, substituting it into the formula (\ref{WKBformula}), one computes $\omega$ for each $\K$ given by (\ref{QNMsK}) with the trial and error way. However, for spherically symmetric black holes, the effective potential depends on the frequency, in most cases, simply as $U(x,\omega)=V(x)-\omega^2$. Therefore we will consider this particular case the next section in more detail.

\section{Spherically symmetric black holes}\label{sec:spherical}

In order to illustrate the main properties of the WKB formula and to test its accuracy, we will consider spherically symmetric black hole, given by the line element
\begin{equation}\label{lineelement}
ds^2=-N(r)dt^2+\frac{B^2(r)}{N(r)}dr^2+r^{D-2}d\Omega_{D-2},
\end{equation}
where $\Omega_{D-2}$ is the line element on a unit $(D-2)$-sphere. The tortoise coordinate is defined as
\begin{equation}\label{tortoise}
  dx=\frac{B(r)}{N(r)}dr,
\end{equation}
so that $x\to\infty$ corresponds to spatial infinity or de Sitter horizon and $x\to-\infty$ corresponds to the event horizon of the black hole.

Any type of linear perturbations, either gravitational or of test fields of various spin and mass propagating in the black-hole background, can be represented as a superposition of multipoles on the $(D-2)$-sphere, whose dynamics can be reduced to a set of wavelike equations. If the final set of equations can be decoupled, the effective potential usually has the form\footnote{The above WKB approach for a single wave-like equation can be modified to work with a set of coupled wavelike equations as well (see e.~g. \cite{Yabana,Bennet}).}
\begin{equation}\label{ssep}
  U(x,\omega)=V(x)-\omega^2,
\end{equation}
where
\begin{eqnarray}
  \lim_{x\to-\infty}V(x) &=& 0, \\
  \lim_{x\to\infty}V(x) &=& \mu^2,
\end{eqnarray}
where the constant $\mu^2$ is square of the field mass, corresponding to the usual dispersion relation for the field far from the black hole,
\begin{equation}\label{sskp}
\omega^2=k_+^2+\mu^2,
\end{equation}
and
\begin{equation}\label{sskm}
k_-^2=\omega^2.
\end{equation}
In order to have an agreement with the initial definition of the ingoing (\ref{ingoingwave}) and outgoing (\ref{outgoingwave}) waves, we define $k_\pm$ in (\ref{sskp}) and (\ref{sskm}) in such a way that $\re{k_\pm}>0$.

It is worth noting that the effective potential $U(x,\omega)$ cannot be represented in the form (\ref{ssep}), not only when considering axisymmetric background, but also for a charged scalar field in the background of the electrically charged spherically symmetric black hole \cite{Konoplya:2002ky}. Although the WKB formula can be applied for these cases, one should keep in mind that, as a rule, the dispersion relation on the horizon differs from (\ref{sskm}). The WKB formula (\ref{WKBformula}) remains valid only when $\re{k_-}$ and $\re{\omega}$ are of the same sign. That is why, for instance, the WKB formula needs modifications when studying phenomenon of superradiance (see Sec.~\ref{sec:using}).

Since derivatives of the potential (\ref{ssep}) do not depend on $\omega$, we observe that, in spherically symmetric background, the WKB formula (\ref{WKBformula}) provides a closed form for the quasinormal frequencies,
\begin{eqnarray}\label{WKBformula-spherical}
\omega^2&=&V_0+A_2(\K^2)+A_4(\K^2)+A_6(\K^2)+\ldots\\\nonumber&-&\imo \K\sqrt{-2V_2}\left(1+A_3(\K^2)+A_5(\K^2)+A_7(\K^2)\ldots\right),
\end{eqnarray}
where $\K$ takes halfinteger values (\ref{QNMsK}), and $V_0, V_2, V_3\ldots$ are, respectively, the value and higher derivatives of the potential $V(x)$ in the maximum, which appear on the righthand side of (\ref{WKBformula-spherical}) and do not depend on the value of $\omega$.

Note, that increasing of the WKB order does not always lead to a better approximation for the frequency. In practice, for a given potential, there is some order, when the WKB formula (\ref{WKBformula-spherical}) provides the best approximation, and using of higher-order formula increases the error. Usually, two sequential orders are compared, in order to estimate the error of the WKB formula approximation. However, it is easy to see from (\ref{WKBformula-spherical}) that, for the real effective potential $V(x)$, each WKB order correction affects either real or imaginary part of the squared frequencies. That is why, for the error estimation for $\omega_k$, obtained with the WKB formula of the order $k$, we use quantity
\begin{equation}\label{errorestimation}
\Delta_k=\frac{|\omega_{k+1}-\omega_{k-1}|}{2}.
\end{equation}
In Sec.~\ref{sec:massivescalar} we show that $\Delta_k$ provides a very good estimation of the error order, usually satisfying,
$$\Delta_k\gtrsim|\omega-\omega_k|,$$
where $\omega$ is the accurate value of the quasinormal frequency.

In order to compare results for various potentials and modes, we also use relative error of the WKB formula of the order $k$,
\begin{equation}\label{relerror}
E_k=\left|\frac{\omega_k-\omega}{\omega}\right|\times 100\%.
\end{equation}

\section{Further improvement of the WKB formula accuracy}\label{sec:Pade}

In order to increase accuracy of the higher-order WKB formula, it has been recently proposed to use Padé approximants \cite{PadeApproximation} for the usual WKB formula \cite{Matyjasek:2017psv}. Within this approach one has to define a polynomial $P_k(\epsilon)$ by introducing powers of the order parameter $\epsilon$ in the righthand side of the WKB formula (\ref{WKBformula-spherical}) as follows,
\begin{eqnarray}\nonumber
  P_k(\epsilon)&=&V_0+A_2(\K^2)\epsilon^2+A_4(\K^2)\epsilon^4+A_6(\K^2)\epsilon^6+\ldots\\&-&\imo \K\sqrt{-2V_2}\left(\epsilon+A_3(\K^2)\epsilon^3+A_5(\K^2)\epsilon^5\ldots\right),\label{WKBpoly}
\end{eqnarray}
where the polynomial order $k$ coincides with the order of the WKB formula. Formal parameter $\epsilon$ is introduced in the same way as in \cite{Iyer:1986np} in order to keep
track of orders in the WKB approximation, and the squared frequency can be obtained by taking $\epsilon=1$,
$$\omega^2=P_k(1).$$

We consider a family of the Padé approximants $P_{\tilde{n}/\tilde{m}}(\epsilon)$ for the polynomial $P_k(\epsilon)$ near $\epsilon=0$ with $\tilde{n}+\tilde{m}=k$, i.~e. we construct rational functions
\begin{equation}\label{WKBPade}
P_{\tilde{n}/\tilde{m}}(\epsilon)=\frac{Q_0+Q_1\epsilon+\ldots+Q_{\tilde{n}}\epsilon^{\tilde{n}}}{R_0+R_1\epsilon+\ldots+R_{\tilde{m}}\epsilon^{\tilde{m}}},
\end{equation}
such that
$$P_{\tilde{n}/\tilde{m}}(\epsilon)-P_k(\epsilon)={\cal O}\left(\epsilon^{k+1}\right).$$

The latter yields that the representation (\ref{WKBPade}) is equivalent to (\ref{WKBpoly}) up to the given order $k$, i.~e. the coefficients $Q_0,Q_1,\ldots,Q_{\tilde{n}}$ and $R_0,R_1,\ldots,R_{\tilde{m}}$ can be obtained in the same way as $A_2,A_3,\ldots,A_k$ by matching the expansion of the solution near the potential peak through the turning points. However, it is more convenient to calculate the coefficients $Q_0,Q_1,\ldots$ and $R_0,R_1,\ldots$ numerically, once the righthand side of (\ref{WKBformula-spherical}) is known.
Finally, the rational function $P_{\tilde{n}/\tilde{m}}(\epsilon)$ is used to approximate squared frequency,
\begin{equation}\label{omegaPade}
\omega^2=P_{\tilde{n}/\tilde{m}}(1).
\end{equation}

For instance, the eikonal formula
\begin{equation}\label{eikonalformula}
\omega^2=P_1(1)=P_{1/0}(1)=V_0-\imo \K\sqrt{-2V_2},
\end{equation}
can be transformed to the following form,
\begin{equation}\label{eikonalalt}
\omega^2=P_{0/1}(1)=\frac{V_0^2}{V_0+\imo \K\sqrt{-2V_2}}.
\end{equation}
In a similar manner, the WKB formula of order $k$, $$\omega^2=P_k(1),$$ can be transformed to the alternative forms, corresponding to $\tilde{m}=1,2,\ldots,k$. We will refer to these expressions as Padé approximations of the order $k$.

It is clear that, when the higher-order corrections are sufficiently small, all the alternative approximations give similar results. However, it turns out, that, in practice, when $\tilde{n}\approx\tilde{m}$, the approximation is much better comparing to the initial WKB formula. In \cite{Matyjasek:2017psv}, $P_{6/6}(1)$ and $P_{6/7}(1)$ were compared to the 6th-order WKB formula $P_{6/0}(1)$. We observed that usually even $P_{3/3}(1)$, i.~e. a Padé approximation of the 6th-order, gives a more accurate value for the squared frequency than $P_{6/0}(1)$ (see Sec.~\ref{sec:massivescalar}).

This approach, being introduced to the particular form of the potential (\ref{ssep}), cannot be straightforwardly generalized for any $U(x,\omega)$ because the transformation is applied to the righthand side after the particular choice of the lefthand side in (\ref{WKBformula-spherical}). Indeed, if we add some constant to both sides of (\ref{WKBformula-spherical}) and repeat the procedure, we obtain different formulas for $\omega^2$. Generally, we observe that such formulas give worse approximations comparing to the case(\ref{omegaPade}) when the lefthand side of is precisely $\omega^2$. This suggests, that the simplest way to use the above improvement for the general formula (\ref{WKBformula}) is adding $\omega^2$ to both sides and replacing the polynomial on the righthand side by the corresponding Padé approximants. Nevertheless, it is possible, that the best accuracy can be achieved with a more sophisticated procedure, e.~g., we can substitute $U(x,\omega/\epsilon)$ into (\ref{WKBformula}) and use Padé approximants on the righthand side. We believe that the best approach for the general potential can be formulated after understanding why we have better accuracy improvement when Padé approximants are used for $\omega^2$ and not for some other quantity. However, we leave these questions to future studies.

\section{Restrictions on using the WKB formula}\label{sec:using}

Summarizing the above sections, the WKB method provides quite a simple and powerful tool for studying properties of black holes. It can be used for solving the scattering problem, which is necessary to find grey-body factors of the black hole, and for calculation of quasinormal modes. In the next section we shall see that the Padé approximation improves accuracy of the WKB approach, allowing one to \emph{calculate the astrophysically relevant quasinormal modes with practically sufficient precision}.

Yet, the downside of simplicity of the approach is that the assumptions made for deducing the WKB formula impose strict limits on the range of its applicability. In particular, the WKB formula (\ref{WKBformula}) cannot be applied when studying:
\begin{enumerate}
\item \textbf{Superradiance.} Owing to extracting rotational energy from a black hole, the incident wave can be reflected with larger amplitude than it had in the beginning. This phenomenon, when the reflection coefficient $|R|$ can be larger than unity is called superradiance \cite{Starobinsky:1973aij}. It can also occur with the charged scalar field in the non-rotating electrically charged black-hole background \cite{Bekenstein:1973mi}. It easy to see from (\ref{reflection}) that the reflection coefficient cannot be larger than unity for any imaginary $\K$, implying that superradiance cannot be described by the WKB formula at least when $U(x,\omega)$ is real. The reason is the specific boundary condition at the horizon, which was not taken into account in (\ref{reflection}). Namely, we choose
     \begin{equation}\label{superradiance}
     k_-=\omega-\omega_s,
     \end{equation}
     where $\omega_s$ is a constant which defines the superradiant regime (see e.~g.~\cite{Brito:2015oca} for review).
     In this way, in the superradiance regime ($\omega<\omega_s$) the group and phase velocities of the field have opposite signs, yielding energy extraction from a black hole. However, the choice of $k_-$ (\ref{superradiance}) implies that in the superradiance regime $k_-<0$, being inconsistent with the initial definition of the ingoing wave. Thus, the WKB formula (\ref{WKBformula}) needs modifications  in order to describe the regime of $\omega<\omega_s$ correctly.
\item \textbf{Stability.} It is easy to see in spherical symmetry from (\ref{WKBformula-spherical}) that, when $A_3, A_5, A_7\ldots$ describe only corrections to the eikonal value, the choice of $\K$ in (\ref{QNMsK}) always leads to $\im{\omega}<0$, corresponding to decaying oscillations. The main reason for that is the assumption that the boundary conditions are a combination of the ingoing and outgoing waves. However, unstable modes correspond to the bound states, and the analytic continuation in this case should be done in a different way.
\item \textbf{Infinitely long-lived modes.}  The same reason is why the WKB formula cannot be used for calculation of \emph{quasiresonances}, which are arbitrarily long lived modes \cite{Ohashi:2004wr}. These modes correspond to solutions with almost zero amplitudes in the asymptotic regions \cite{Konoplya:2004wg} and, therefore, cannot be adequately described by the WKB formula. Practically, the WKB approach allows one to calculate frequencies of massive fields close to the quasiresonance regime, when the multipole number $\ell$ is large. Yet, for large values of the field mass, the effective potential does not have a local maximum \cite{Ohashi:2004wr}, so that the WKB expansion cannot be performed.
\item \textbf{Higher overtones of the quasinormal spectrum.} The analytic continuation of (\ref{WKBformula}) in the complex plane works well only for $|\re{\omega}|\gtrsim|\im{\omega}|$, providing bad approximation for the modes with high decay rate. Usually the WKB accuracy is reasonable for $\ell > n$ and marginal already for $\ell=n$. However, in the next section we shall see that usage of the Pade approximants can considerably improve the accuracy allowing one find even several first overtones $n> \ell$.
\item \textbf{Asymptotically nonconstant potential.} Some of the asymptotically non-constant effective potentials require qualitatively different boundary conditions, e.~g. black holes in the anti-de Sitter universe usually require Dirichlet boundary conditions at infinity. In that case there is no sense in the WKB-expansion in the  form (\ref{WKBformula}). However, in a number of particular cases, for example, when studying holographic superconductors  \cite{Hartnoll:2008kx,Konoplya:2009hv}, WKB formula can be applied for the asymptotically anti-de Sitter geometry, because the effective potential vanishes at infinity.
\end{enumerate}

\section{Higher dimensions and massive fields: strategy for the choice of WKB order and Padé approximants}\label{sec:massivescalar}
Here we will study perturbations of the Tangherlini black hole \cite{Tangherlini:1963bw} of unit radius, given by the metric (\ref{lineelement}) with
$$N(r)=1-r^{3-D},\qquad B(r)=1.$$
The radial part of the Klein-Gordon equation for the field of mass $\mu$ is reduced to the wavelike equation (\ref{wavelike}) with the effective potential
\begin{eqnarray}\label{test-scalar}
U&=&V(r)-\omega^2=N(r)\Biggr(\mu^2+\frac{\ell(\ell+D-3)}{r^2}\\\nonumber&&+\frac{D-2}{2r}N'(r)+\frac{(D-4)(D-2)}{4r^2}N(r)\Biggr)-\omega^2,
\end{eqnarray}
where $\ell$ is the multipole number and the tortoise coordinate is given by (\ref{tortoise}).

\subsection{Quasinormal modes of massless scalar field}

\begin{table}
\begin{tabular}{|r|c|c|c|r|}
\hline
$k$ & $\omega_k$ & $\Delta_k$ & $|\omega-\omega_k|$ & $E_k$ \\
\hline
\multicolumn{5}{|c|}{$\ell=0$, $n=0$: $\omega=0.220910-0.209791\imo$}\\
\hline
$ 3$ & $0.209294-0.230394\imo$ & $0.038352$ & $0.023651$ & $  7.763\%$ \\
$ 4$ & $0.219199-0.219982\imo$ & $0.009486$ & $0.010334$ & $  3.392\%$ \\
$ 5$ & $0.210655-0.211471\imo$ & $0.009216$ & $0.010391$ & $  3.411\%$ \\
$ 6$ & $0.220934-0.201633\imo$ & $0.007917$ & $0.008159$ & $  2.678\%$ \\
$ 7$ & $0.225845-0.207002\imo$ & $0.005886$ & $0.005669$ & $  1.861\%$ \\
$ 8$ & $0.232684-0.200917\imo$ & $0.018639$ & $0.014744$ & $  4.839\%$ \\
$ 9$ & $0.256553-0.228135\imo$ & $0.028661$ & $0.040087$ & $ 13.158\%$ \\
$10$ & $0.226902-0.257947\imo$ & $0.041922$ & $0.048527$ & $ 15.929\%$ \\
$11$ & $0.282397-0.307896\imo$ & $0.180081$ & $0.115781$ & $ 38.004\%$ \\
\hline
\multicolumn{5}{|c|}{$\ell=0$, $n=1$: $\omega=0.172234-0.696105\imo$}\\
\hline
$ 3$ & $0.178378-0.709920\imo$ & $0.063122$ & $0.015120$ & $ 2.108\%$ \\
$ 4$ & $0.177166-0.714777\imo$ & $0.003676$ & $0.019312$ & $ 2.693\%$ \\
$ 5$ & $0.171958-0.713504\imo$ & $0.012867$ & $0.017401$ & $ 2.427\%$ \\
$ 6$ & $0.178058-0.689058\imo$ & $0.013437$ & $0.009143$ & $ 1.275\%$ \\
$ 7$ & $0.169056-0.686786\imo$ & $0.005143$ & $0.009846$ & $ 1.373\%$ \\
$ 8$ & $0.170111-0.682528\imo$ & $0.022665$ & $0.013742$ & $ 1.916\%$ \\
$ 9$ & $0.213694-0.694674\imo$ & $0.023072$ & $0.041485$ & $ 5.785\%$ \\
$10$ & $0.210747-0.704390\imo$ & $0.070391$ & $0.039394$ & $ 5.494\%$ \\
$11$ & $0.341425-0.753872\imo$ & $0.209742$ & $0.178781$ & $24.931\%$ \\
\hline
\end{tabular}
\caption{Quasinormal modes of the massless scalar field for $D=4$, $\ell=0$ calculated with the WKB formula of different orders: The error estimation $\Delta_k=|\omega_{k+1}-\omega_{k-1}|/2$ allows to determine the two dominant modes with accuracy of less than $2\%$ with the help of the WKB formula of 7th-order.}\label{tabl:masslessl0}
\end{table}

\begin{table}
\begin{tabular}{|r|c|c|c|r|}
\hline
$k$ & $\omega_k$ & $\Delta_k$ & $|\omega-\omega_k|$ & $E_k$ \\
\hline
\multicolumn{5}{|c|}{$\ell=1$, $n=0$: $\omega=0.585872-0.195320\imo$}\\
\hline
$ 2$ & $0.589020-0.215341\imo$ & $0.038360$ & $0.020267$ & $3.282\%$ \\
$ 3$ & $0.582228-0.196003\imo$ & $0.010402$ & $0.003707$ & $0.600\%$ \\
$ 4$ & $0.585907-0.194772\imo$ & $0.001969$ & $0.000549$ & $0.089\%$ \\
$ 5$ & $0.586124-0.195422\imo$ & $0.000378$ & $0.000271$ & $0.044\%$ \\
$ 6$ & $0.585819-0.195523\imo$ & $0.000197$ & $0.000210$ & $0.034\%$ \\
$ 7$ & $0.585746-0.195305\imo$ & $0.000130$ & $0.000127$ & $0.021\%$ \\
$ 8$ & $0.585862-0.195267\imo$ & $0.000061$ & $0.000054$ & $0.009\%$ \\
$ 9$ & $0.585867-0.195284\imo$ & $0.000040$ & $0.000036$ & $0.006\%$ \\
$10$ & $0.585942-0.195259\imo$ & $0.000104$ & $0.000093$ & $0.015\%$ \\
$11$ & $0.586003-0.195441\imo$ & $0.000200$ & $0.000178$ & $0.029\%$ \\
$12$ & $0.585669-0.195552\imo$ & $0.000349$ & $0.000308$ & $0.050\%$ \\
\hline
\multicolumn{5}{|c|}{$\ell=1$, $n=1$: $\omega=0.528897-0.612515\imo$}\\
\hline
$ 2$ & $0.576605-0.659933\imo$ & $0.149886$ & $0.067265$ & $8.312\%$ \\
$ 3$ & $0.524424-0.614865\imo$ & $0.034551$ & $0.005053$ & $0.624\%$ \\
$ 4$ & $0.527253-0.611565\imo$ & $0.002442$ & $0.001899$ & $0.235\%$ \\
$ 5$ & $0.528949-0.613028\imo$ & $0.001120$ & $0.000516$ & $0.064\%$ \\
$ 6$ & $0.528942-0.613036\imo$ & $0.000204$ & $0.000523$ & $0.065\%$ \\
$ 7$ & $0.528634-0.612771\imo$ & $0.000252$ & $0.000367$ & $0.045\%$ \\
$ 8$ & $0.528827-0.612547\imo$ & $0.000192$ & $0.000077$ & $0.009\%$ \\
$ 9$ & $0.528642-0.612387\imo$ & $0.000312$ & $0.000285$ & $0.035\%$ \\
$10$ & $0.529018-0.611952\imo$ & $0.000636$ & $0.000576$ & $0.071\%$ \\
$11$ & $0.529876-0.612694\imo$ & $0.001118$ & $0.000995$ & $0.123\%$ \\
$12$ & $0.528618-0.614152\imo$ & $0.001861$ & $0.001661$ & $0.205\%$ \\
\hline
\multicolumn{5}{|c|}{$\ell=1$, $n=2$: $\omega=0.459079-1.080267\imo$}\\
\hline
$ 2$ & $0.570510-1.11164\imo$ & $0.304088$ & $0.115763$ & $9.863\%$ \\
$ 3$ & $0.447086-1.05363\imo$ & $0.069325$ & $0.029208$ & $2.488\%$ \\
$ 4$ & $0.434729-1.08358\imo$ & $0.020575$ & $0.024575$ & $2.094\%$ \\
$ 5$ & $0.458264-1.09324\imo$ & $0.013655$ & $0.012996$ & $1.107\%$ \\
$ 6$ & $0.462028-1.08433\imo$ & $0.004874$ & $0.005021$ & $0.428\%$ \\
$ 7$ & $0.463192-1.08483\imo$ & $0.004308$ & $0.006141$ & $0.523\%$ \\
$ 8$ & $0.466558-1.07700\imo$ & $0.006996$ & $0.008161$ & $0.695\%$ \\
$ 9$ & $0.456354-1.07262\imo$ & $0.006482$ & $0.008118$ & $0.692\%$ \\
$10$ & $0.453724-1.07884\imo$ & $0.004678$ & $0.005543$ & $0.472\%$ \\
$11$ & $0.459693-1.08136\imo$ & $0.004306$ & $0.001255$ & $0.107\%$ \\
$12$ & $0.457483-1.08658\imo$ & $0.003013$ & $0.006516$ & $0.555\%$ \\
\hline
\multicolumn{5}{|c|}{$\ell=1$, $n=3$: $\omega=0.406517-1.576596\imo$}\\
\hline
$ 3$ & $0.347404-1.49726\imo$ & $0.127524$ & $0.098938$ & $ 6.077\%$ \\
$ 4$ & $0.320285-1.62404\imo$ & $0.085916$ & $0.098421$ & $ 6.045\%$ \\
$ 5$ & $0.428444-1.64878\imo$ & $0.064202$ & $0.075441$ & $ 4.634\%$ \\
$ 6$ & $0.444189-1.59034\imo$ & $0.031408$ & $0.040100$ & $ 2.463\%$ \\
$ 7$ & $0.461173-1.59516\imo$ & $0.057908$ & $0.057724$ & $ 3.545\%$ \\
$ 8$ & $0.494969-1.48625\imo$ & $0.117273$ & $0.126438$ & $ 7.766\%$ \\
$ 9$ & $0.292972-1.43170\imo$ & $0.144544$ & $0.184084$ & $11.306\%$ \\
$10$ & $0.254679-1.64697\imo$ & $0.127054$ & $0.167355$ & $10.279\%$ \\
$11$ & $0.379148-1.67075\imo$ & $0.076763$ & $0.098052$ & $ 6.022\%$ \\
$12$ & $0.360239-1.75845\imo$ & $0.179682$ & $0.187649$ & $11.525\%$ \\
\hline
\end{tabular}
\caption{Quasinormal modes of the massless scalar field for $D=4$, $\ell=1$ calculated with the WKB formula of different orders. The error estimation $\Delta_k=|\omega_{k+1}-\omega_{k-1}|/2$ allows to determine the WKB order in which the error is minimal.}\label{tabl:masslessl1}
\end{table}

We start from the massless ($\mu=0$) field in $D=4$ dimensions. It is well known that the WKB formula gives very accurate frequencies for high $\ell$ and small overtone numbers. In the tables~\ref{tabl:masslessl0}~and~\ref{tabl:masslessl1} we summarize accuracy of the WKB formula for $\ell=0$ and $\ell=1$. We see that the error can be very well estimated by comparing frequencies, obtained with the help of the WKB formula at different orders. The quantity $\Delta_k$, introduced in (\ref{errorestimation}), not only allows one to find the order of the absolute error but also to determine the order, which gives the most accurate approximation for the quasinormal mode.

\begin{table}
\begin{tabular}{|r|c|c|c|r|}
\hline
$k$ & $\omega_k$ & $\Delta_k$ & $|\omega-\omega_k|$ & $E_k$ \\
\hline
\multicolumn{5}{|c|}{$\ell=0$, $n=0$: $\omega=1.270541-0.665778\imo$}\\
\hline
$2$ & $1.30195-0.918489\imo$ & $0.282472$ & $0.254656$ & $17.8\%$ \\
$3$ & $1.14476-0.677518\imo$ & $0.154943$ & $0.126326$ & $ 8.8\%$ \\
$4$ & $1.27119-0.610132\imo$ & $0.075412$ & $0.055649$ & $ 3.9\%$ \\
$5$ & $1.29412-0.656568\imo$ & $0.031410$ & $0.025317$ & $ 1.8\%$ \\
$6$ & $1.32604-0.640766\imo$ & $0.109626$ & $0.060873$ & $ 4.2\%$ \\
$7$ & $1.42748-0.830601\imo$ & $0.356862$ & $0.227587$ & $15.9\%$ \\
$8$ & $0.950349-1.24761\imo$ & $1.035640$ & $0.664117$ & $46.3\%$ \\
\hline
\multicolumn{5}{|c|}{$\ell=0$, $n=1$: $\omega=0.683428-2.438793\imo$}\\
\hline
$2$ & $1.296960-2.76607\imo$ & $0.931223$ & $0.695363$ & $27.5\%$ \\
$3$ & $0.615189-2.51942\imo$ & $0.363277$ & $0.105627$ & $ 4.2\%$ \\
$4$ & $0.618369-2.50646\imo$ & $0.047304$ & $0.093871$ & $ 3.7\%$ \\
$5$ & $0.526880-2.48547\imo$ & $0.117573$ & $0.163359$ & $ 6.5\%$ \\
$6$ & $0.575563-2.27524\imo$ & $0.287178$ & $0.195915$ & $ 7.7\%$ \\
$7$ & $1.100720-2.46111\imo$ & $0.361402$ & $0.417888$ & $16.5\%$ \\
$8$ & $0.933017-2.90347\imo$ & $0.547973$ & $0.527468$ & $20.8\%$ \\
\hline
\end{tabular}
\caption{Quasinormal modes of the massless scalar field for $D=7$, $\ell=0$ calculated with the WKB formula of different orders: The error estimation $\Delta_k=|\omega_{k+1}-\omega_{k-1}|/2$ is worse when the potential has a complex form, yet still gives a correct order of the error.}\label{tabl:masslessl0D7}
\end{table}

When the number of space-time dimensions $D$ increases, the accuracy of the higher-order WKB formula falls down quickly (see table~\ref{tabl:masslessl0D7}), which makes the higher orders virtually not useful without the Padé approximation described in Sec.~\ref{sec:Pade}.

\begin{figure*}
\resizebox{\linewidth}{!}{\includegraphics*{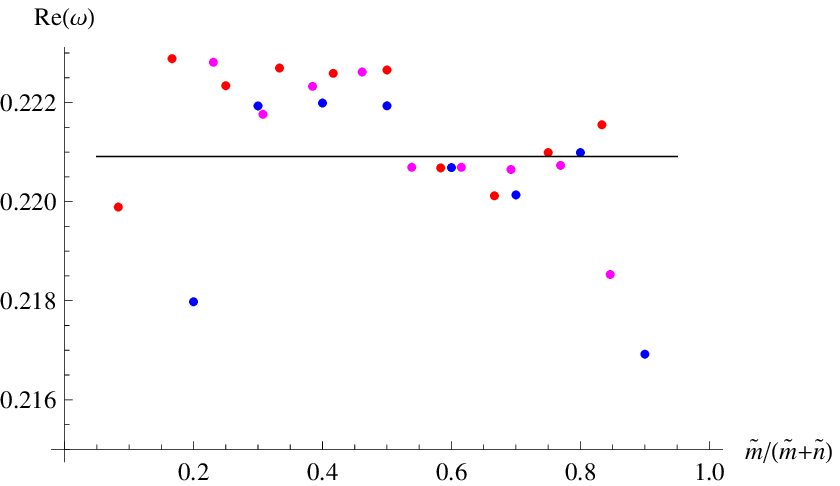}\includegraphics*{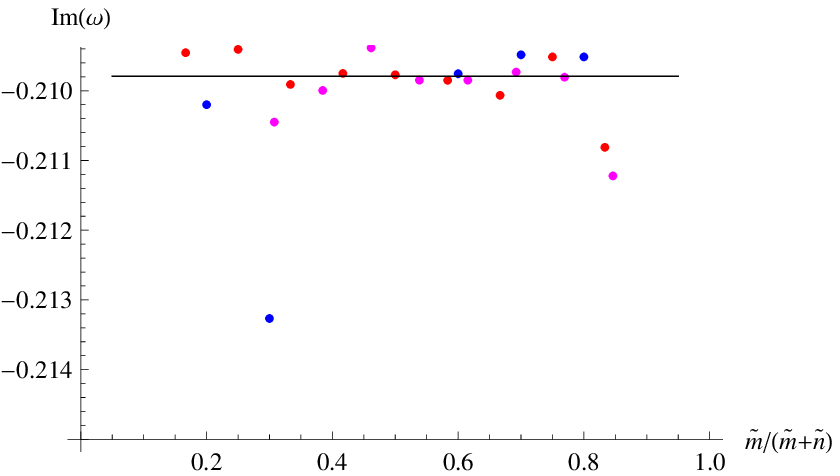}}
\caption{Real (left panel) and imaginary (right panel) parts of the dominant frequency ($D=4$, $\ell=0$, $n=0$) of massless scalar field ($\mu=0$) calculated using Padé approximants, $P_{\tilde{n}/\tilde{m}}$, for the WKB formula of the orders $\tilde{n}+\tilde{m}=10$ (blue), $\tilde{n}+\tilde{m}=12$ (red), and $\tilde{n}+\tilde{m}=13$ (magenta). Solid line corresponds to the accurate value of the frequency.}\label{fig:Pade}
\end{figure*}

On Fig.~\ref{fig:Pade} we plot the fundamental quasinormal frequency ($D=4$, $\ell=0$, $n=0$) obtained with the help of various Padé approximants $P_{\tilde{n}/\tilde{m}}$. One can see that the best accuracy, within a given order $k$, is achieved when $\tilde{n}\approx\tilde{m}\approx k/2$. In \cite{Matyjasek:2017psv} such kind of the approximants, $P_{6/6}$ and $P_{6/7}$ of twelfth and thirteenth orders WKB, were studied, showing much better accuracy than the sixth order the usual WKB formula \cite{Konoplya:2003ii} for a number of cases. Here we will consider possible variants of choice of parameters for the Padé approximation at a given WKB order for various configurations of a black hole and field in its vicinity, in order to understand which choice corresponds to the best accuracy.

We notice that indeed good approximations tend to gather near $\tilde{n}\approx\tilde{m}\approx k/2$ in the great majority of cases. Nevertheless, for some values of the parameters the optimal choice of the Padé approximant occurs for $\tilde{n}$ which is much different from $\tilde{m}$. In some cases, usually when either $\tilde{n}$ or $\tilde{m}$ is small, Padé approximations lead to values, which lay far from the accurate value of the quasinormal frequency and, at the same time, far from the most of the values given by other Padé approximations at the same WKB order.

In order to illustrate this observation we shall use quasinormal frequencies obtained with the help of the Padé approximation given by (\ref{omegaPade}) at the ($k=\tilde{n}+\tilde{m}$)-WKB order and designate this frequency as $\omega_{\tilde{n}/\tilde{m}}$. Then, using $\omega_{\tilde{n}/\tilde{m}}$ we will construct the following quantities:
\begin{enumerate}
\item The central value
$$\omega_k^{(c)}=\omega_{[k/2]/[(k+1)/2]},$$
which corresponds to $\omega_{1/1}$, $\omega_{1/2}$, $\omega_{2/2}$, $\omega_{2/3}$, $\omega_{3/3}$, $\omega_{3/4}$, $\omega_{4/4}$, $\omega_{4/5}$, $\omega_{5/5}$, $\omega_{5/6}$, $\omega_{6/6}$, and $\omega_{6/7}$, respectively, for each $WKB$ order from 2-nd to 13-th.
\item Mean value for the two (for odd $k$) or three (for even $k$) central values, $\omega_k^{(m)}$, defined as
$$\omega_k^{(m)}=\left\{
                        \begin{array}{ll}
                          \displaystyle\frac{1}{3}\sum_{\tilde{m}=p-1}^{p+1}\omega_{(k-\tilde{m})/\tilde{m}}, & k=2p, \\
                          \displaystyle\frac{1}{2}\sum_{\tilde{m}=p}^{p+1}\omega_{(k-\tilde{m})/\tilde{m}}, & k=2p+1.
                        \end{array}
                      \right.
$$
In this way, for the eikonal formula we take mean of the values, obtained with the formulas~(\ref{eikonalformula})~and~(\ref{eikonalalt}); for the second-order approximation we take mean of all three frequencies, and for higher orders we take into account only the central values, for which $\tilde{m}=\tilde{n}\pm1$ (odd order) or $\tilde{m}=\tilde{n}$ and $\tilde{m}=\tilde{n}\pm2$ (even order).
\item The mean value of two closest frequencies, corresponding to sequent values of $\tilde{m}$ and $\tilde{n}$
$$\omega_k^{(1)}=\frac{1}{2}\left(\omega_{(\tilde{n}+1)/(\tilde{m}-1)}+\omega_{\tilde{n}/\tilde{m}}\right),$$
i.~e., we choose such value of $\tilde{m}$, for which the relative difference of the above frequencies is minimal.
\item The mean value of the second pair of closest frequencies, $\omega_k^{(2)}$.
\end{enumerate}

\begin{table}
\begin{tabular}{|r|r|r|r|r|r|r|}
\hline
 $k$ & $\omega_{k/0}$ & $\omega_k^{(c)}$ & $\omega_k^{(m)}$ & $\omega_k^{(1)}$ & $\omega_k^{(2)}$ \\
\hline
 $2$ & $  17.8\%$ & $3.934\%$ & $5.230\%$ & $1.887\%$ & $8.102\%$ \\
 $3$ & $   8.8\%$ & $0.555\%$ & $0.672\%$ & $0.537\%$ & $0.672\%$ \\
 $4$ & $   3.9\%$ & $0.402\%$ & $0.222\%$ & $0.218\%$ & $0.271\%$ \\
 $5$ & $   1.9\%$ & $0.590\%$ & $0.626\%$ & $0.566\%$ & $0.626\%$ \\
 $6$ & $   4.2\%$ & $0.419\%$ & $0.250\%$ & $0.372\%$ & $0.158\%$ \\
 $7$ & $  15.9\%$ & $0.034\%$ & $0.251\%$ & $0.034\%$ & $0.522\%$ \\
 $8$ & $  46.3\%$ & $0.114\%$ & $0.118\%$ & $0.068\%$ & $0.070\%$ \\
 $9$ & $ 130.4\%$ & $0.026\%$ & $0.110\%$ & $0.026\%$ & $0.218\%$ \\
$10$ & $  90.2\%$ & $0.009\%$ & $0.109\%$ & $0.013\%$ & $0.012\%$ \\
$11$ & $ 360.3\%$ & $0.023\%$ & $0.059\%$ & $0.023\%$ & $0.108\%$ \\
$12$ & $ 951.1\%$ & $0.007\%$ & $0.143\%$ & $0.015\%$ & $0.015\%$ \\
$13$ & $2032.6\%$ & $0.011\%$ & $0.011\%$ & $0.011\%$ & $0.011\%$ \\
\hline
\end{tabular}
\caption{Relative errors for the dominant mode of the massless scalar field ($\ell=0$, $n=0$) in $D=7$, obtained within the same WKB order using different Padé approximants. The accurate value is $\omega=1.27054-0.66578\imo$.}\label{tabl:WKBerror0}
\end{table}

From the table~\ref{tabl:WKBerror0} we see that although for the dominant mode $\omega_k^{(c)}$ always provides better approximation than the ordinary WKB formula, and, usually, than $\omega_k^{(m)}$, \emph{the best approximation is given by} $\omega_k^{(1)}$.

\begin{table}
\begin{tabular}{|r|r|r|r|r|r|r|}
\hline
 $k$ & $\omega_{k/0}$ & $\omega_k^{(c)}$ & $\omega_k^{(m)}$ & $\omega_k^{(1)}$ & $\omega_k^{(2)}$ \\
\hline
 $2$ & $  27.5\%$ & $40.51\%$ & $32.5\%$ & $22.73\%$ & $54.37\%$ \\
 $3$ & $   4.2\%$ & $38.52\%$ & $21.8\%$ & $ 2.86\%$ & $21.84\%$ \\
 $4$ & $   3.7\%$ & $11.18\%$ & $56.4\%$ & $ 3.70\%$ & $ 5.70\%$ \\
 $5$ & $   6.5\%$ & $14.91\%$ & $10.5\%$ & $ 5.12\%$ & $ 4.78\%$ \\
 $6$ & $   7.7\%$ & $ 5.93\%$ & $16.0\%$ & $ 3.01\%$ & $ 3.80\%$ \\
 $7$ & $  16.5\%$ & $37.33\%$ & $18.1\%$ & $ 3.60\%$ & $ 3.60\%$ \\
 $8$ & $  20.8\%$ & $30.51\%$ & $21.0\%$ & $ 3.01\%$ & $31.70\%$ \\
 $9$ & $  55.9\%$ & $37.48\%$ & $17.7\%$ & $ 4.96\%$ & $ 5.34\%$ \\
$10$ & $ 229.0\%$ & $11.27\%$ & $18.1\%$ & $ 3.54\%$ & $ 4.25\%$ \\
$11$ & $ 693.1\%$ & $15.51\%$ & $ 7.8\%$ & $ 3.47\%$ & $ 3.49\%$ \\
$12$ & $ 964.6\%$ & $15.90\%$ & $10.4\%$ & $ 3.48\%$ & $15.97\%$ \\
$13$ & $1430.1\%$ & $15.48\%$ & $ 9.2\%$ & $ 5.71\%$ & $ 4.50\%$ \\
\hline
\end{tabular}
\caption{Relative errors for the first overtone of the massless scalar field ($\ell=0$, $n=1$) in $D=7$, obtained within the same WKB order using different Padé approximants. The accurate value is $\omega=0.68343-2.43879\imo$.}\label{tabl:WKBerror1}
\end{table}

The picture is more clear if we study the first overtone (see table~\ref{tabl:WKBerror1}). In this case the central values $\omega_k^{(c)}$ in 12th and 13th orders \cite{Matyjasek:2017psv}, give worse approximation than the ordinary WKB formula. However, the better approximation is again achieved by $\omega_k^{(1)}$, even within the same $WKB$ order ($k=6$). We notice also, that $\omega_k^{(2)}$ sometimes provides even better approximation than $\omega_k^{(1)}$, which means that the minimal difference cannot be taken as a universal criterium for choosing the best approximation. Moreover, if one takes the mean value of $\omega_k^{(1)}$ and $\omega_k^{(2)}$, then such an approximation is usually better for larger $k$.

Thus, Padé approximants give better accuracy comparing to the ordinary WKB formula at the same order, what drastically improves the higher-order WKB approximation. However, we do not have a mathematically strict criterium for choosing the appropriate orders $\tilde{n}$ and $\tilde{m}$, for which the approximation is the best or sufficiently close to the best one. Our empirical observation is that we have to exclude somehow the nonsense values, which appear at large distance from the others. The remaining values, laying close to each other, are spread around the accurate value of the frequency. The higher WKB order we consider, the larger number of such approximate values we can find, however, these values do not always correspond to $\tilde{m}\approx\tilde{n}\approx k/2$. As a rule, for the dominant frequencies, the best approximation within the same order corresponds to larger $\tilde{m}$, while for higher overtones smaller values of $\tilde{m}$ provide a better accuracy.

In order to select points, which appear around the accurate value of the frequency, we consider $r$ pairs of the closest values within each order $k$, where $r=[(k+1)/3]$; i.~e., we take $\omega_k^{(1)}$ for $k=1,2,3$; $\omega_k^{(1)}$ and $\omega_k^{(2)}$ for $k=4,5,6$; $\omega_k^{(1)}$, $\omega_k^{(2)}$, and $\omega_k^{(3)}$ for $k=7,8,9$; and so on. We calculate the average of these frequencies, $\omega_k$, as an approximation of the order $k$ and estimate the order of the error with the help of the standard deviation formula for all the values, $\omega_{\tilde{m}/\tilde{n}}$ which we have used for averaging.

\begin{table}
\begin{tabular}{|r|c|c|c|r|}
\hline
$k$ & $\omega_k$ & $S$ & $|\omega-\omega_k|$ & $E_k$ \\
\hline
\multicolumn{5}{|c|}{$\ell=0$, $n=0$: $\omega=0.220910-0.209791\imo$}\\
\hline
$ 1$ & $0.299357-0.154959\imo$ & $0.090322$ & $0.095711$ & $31.416\%$ \\
$ 2$ & $0.228275-0.184305\imo$ & $0.029568$ & $0.026529$ & $ 8.708\%$ \\
$ 3$ & $0.221995-0.200495\imo$ & $0.004228$ & $0.009359$ & $ 3.072\%$ \\
$ 4$ & $0.219557-0.213437\imo$ & $0.007574$ & $0.003889$ & $ 1.277\%$ \\
$ 5$ & $0.222404-0.207297\imo$ & $0.001538$ & $0.002907$ & $ 0.954\%$ \\
$ 6$ & $0.222620-0.209111\imo$ & $0.001214$ & $0.001841$ & $ 0.604\%$ \\
$ 7$ & $0.223831-0.208727\imo$ & $0.000682$ & $0.003109$ & $ 1.021\%$ \\
$ 8$ & $0.221571-0.209650\imo$ & $0.001682$ & $0.000676$ & $ 0.222\%$ \\
$ 9$ & $0.220627-0.209924\imo$ & $0.001888$ & $0.000313$ & $ 0.103\%$ \\
$10$ & $0.221061-0.209373\imo$ & $0.000805$ & $0.000444$ & $ 0.146\%$ \\
$11$ & $0.221398-0.209668\imo$ & $0.000650$ & $0.000503$ & $ 0.165\%$ \\
$12$ & $0.222069-0.209745\imo$ & $0.001006$ & $0.001160$ & $ 0.381\%$ \\
$13$ & $0.221314-0.209864\imo$ & $0.000835$ & $0.000410$ & $ 0.135\%$ \\
\hline
\multicolumn{5}{|c|}{$\ell=0$, $n=1$: $\omega=0.172234-0.696105\imo$}\\
\hline
$ 1$ & $0.326778-0.258711\imo$ & $0.261240$ & $0.463894$ & $64.691\%$ \\
$ 2$ & $0.255452-0.572242\imo$ & $0.182173$ & $0.149222$ & $20.809\%$ \\
$ 3$ & $0.190421-0.692043\imo$ & $0.021555$ & $0.018635$ & $ 2.599\%$ \\
$ 4$ & $0.172263-0.710489\imo$ & $0.011471$ & $0.014384$ & $ 2.006\%$ \\
$ 5$ & $0.173638-0.712474\imo$ & $0.001965$ & $0.016429$ & $ 2.291\%$ \\
$ 6$ & $0.175050-0.710126\imo$ & $0.007182$ & $0.014301$ & $ 1.994\%$ \\
$ 7$ & $0.172976-0.695820\imo$ & $0.007961$ & $0.000795$ & $ 0.111\%$ \\
$ 8$ & $0.169464-0.685270\imo$ & $0.004811$ & $0.011184$ & $ 1.560\%$ \\
$ 9$ & $0.176306-0.693955\imo$ & $0.007789$ & $0.004605$ & $ 0.642\%$ \\
$10$ & $0.188501-0.700536\imo$ & $0.012902$ & $0.016859$ & $ 2.351\%$ \\
$11$ & $0.173366-0.697739\imo$ & $0.011354$ & $0.001988$ & $ 0.277\%$ \\
$12$ & $0.173682-0.695890\imo$ & $0.001044$ & $0.001464$ & $ 0.204\%$ \\
$13$ & $0.198182-0.781001\imo$ & $0.181544$ & $0.088773$ & $12.379\%$ \\
\hline
\multicolumn{5}{|c|}{$\ell=0$, $n=2$: $\omega=0.151484-1.202157\imo$}\\
\hline
$ 1$ & $0.373493-0.324390\imo$ & $0.373172$ & $0.905408$ & $74.724\%$ \\
$ 2$ & $0.231814-0.843204\imo$ & $0.385030$ & $0.367832$ & $30.358\%$ \\
$ 3$ & $0.137504-1.164517\imo$ & $0.026792$ & $0.040153$ & $ 3.314\%$ \\
$ 4$ & $0.110483-1.270431\imo$ & $0.065853$ & $0.079639$ & $ 6.573\%$ \\
$ 5$ & $0.198941-1.243204\imo$ & $0.061432$ & $0.062746$ & $ 5.179\%$ \\
$ 6$ & $0.156048-1.224905\imo$ & $0.004679$ & $0.023201$ & $ 1.915\%$ \\
$ 7$ & $0.157748-1.206593\imo$ & $0.010339$ & $0.007676$ & $ 0.633\%$ \\
$ 8$ & $0.136614-1.201415\imo$ & $0.006416$ & $0.014888$ & $ 1.229\%$ \\
$ 9$ & $0.140454-1.205179\imo$ & $0.001148$ & $0.011436$ & $ 0.944\%$ \\
$10$ & $0.137352-1.200519\imo$ & $0.006915$ & $0.014227$ & $ 1.174\%$ \\
$11$ & $0.157845-1.195389\imo$ & $0.013468$ & $0.009288$ & $ 0.767\%$ \\
$12$ & $0.138801-1.192967\imo$ & $0.083910$ & $0.015663$ & $ 1.293\%$ \\
$13$ & $0.155009-1.197784\imo$ & $0.023563$ & $0.005616$ & $ 0.464\%$ \\
\hline
\end{tabular}
\caption{Quasinormal modes of the massless scalar field for $D=4$, $\ell=0$ calculated using Padé approximants of different orders. Standard deviation $S$ allows to estimate the error using only values of the same WKB order. Although for $n=2$ the standard deviation in the minimum is one order smaller than the corresponding absolute error, Padé approximants of order 9 still give an estimation for the frequency with the error smaller than $1\%$. For $n=0,1$ the Padé approximants improve accuracy by one order comparing to the ordinary WKB formula (cf.~table~\ref{tabl:masslessl0}). }\label{tabl:masslessl0Pade}
\end{table}

On table~\ref{tabl:masslessl0Pade} we see that the approach described above allows one not only to improve the accuracy of the dominant mode by one order, but also to calculate the first and second overtones with reasonable accuracy for $\ell=0$. The accuracy improvement is even larger for $\ell>0$, and, surprisingly, for $D>4$. The latter is due to an interesting fact, that for large $D$ the best accuracy is achieved by usage of the high WKB orders in combination with the Padé approximation. For the scalar field in $D>7$ \cite{Rostworowski:2006bp}, the best approximation can be found with the help of the WKB expansion of order $13$ \cite{Matyjasek:2017psv}. However, for lower numbers of space-time dimensions $D$, the best approximation is achieved already at lower orders. Thus, for practical purpose, it is efficient to use Padé approximants, corresponding to the lower WKB orders. This can be crucial for cumbersome potentials, when the calculation of higher derivatives is a time-consuming task.

\subsection{Quasinormal modes of massive scalar field}

\begin{figure}
\resizebox{\linewidth}{!}{\includegraphics*{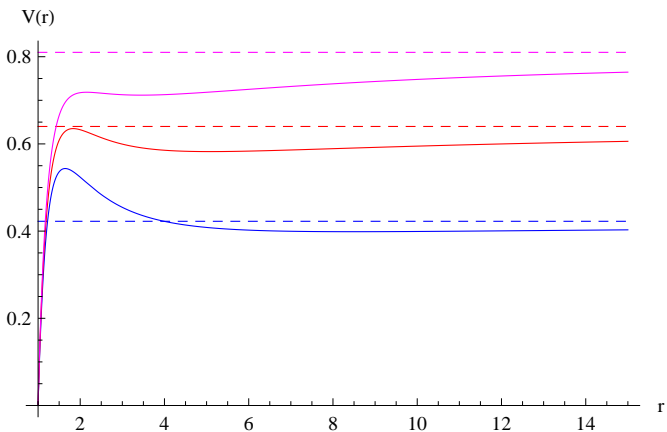}}
\caption{Effective potential for the massive scalar field $\mu=0.65$ (blue, lower), $\mu=0.80$ (red), and $\mu=0.90$ (magenta, upper) for $D=4$, $\ell=1$. The asymptotic value is dotted.}\label{fig:massivescalarpot}
\end{figure}

Although the WKB formula cannot be used to study quasinormal modes in the quasiresonance regime \cite{Konoplya:2017tvu,Zhang:2018jgj}, we observe that the ordinary WKB formula provides a very good approximation even for large mass provided the peak is high enough, i.~e. for $\ell\geq1$. The WKB formula (\ref{WKBformula-spherical}) takes into account only two turning points near the peak of the effective potential. Since the potential (\ref{test-scalar}) has also a minimum to the righthand side of the peak and then grows until reaches the asymptotic value (see fig.~\ref{fig:massivescalarpot}), the accurate approach would require to take into account backscattering from that far barrier. However, as long as the asymptotical value is lower than the peak, neglecting the potential pit does not lead to a significant error.

We compare the accurate values of quasinormal modes of the massive scalar field, $\omega$, found in \cite{Konoplya:2004wg,Zhidenko:2006rs} with the approximation given by the WKB formula at each order, $\omega_k$, and calculate the relative error for the sufficiently large values of the mass, such that $\mu^2\lessapprox V_0$.

\begin{table*}
\begin{tabular}{|l|r|r|r|r|r|r|r|r|r|r|r|r|r|}
\hline
 $\mu$ & 1 order & 2 order & 3 order & 4 order & 5 order & 6 order & 7 order & 8 order & 9 order & 10 order & 11 order & 12 order & 13 order \\
\hline
 $0.65$ & $9.845\%$ & $2.689\%$ & $0.595\%$ & $0.086\%$ & $0.043\%$ & $0.027\%$ & $0.019\%$ & $0.008\%$ & $0.004\%$ & $\bf 0.004\%$ & $0.010\%$ & $0.019\%$ & $0.039\%$ \\
 $0.66$ & $9.819\%$ & $2.663\%$ & $0.599\%$ & $0.084\%$ & $0.044\%$ & $0.027\%$ & $0.020\%$ & $0.007\%$ & $0.005\%$ & $\bf 0.004\%$ & $0.010\%$ & $0.018\%$ & $0.037\%$ \\
 $0.67$ & $9.795\%$ & $2.636\%$ & $0.604\%$ & $0.081\%$ & $0.046\%$ & $0.027\%$ & $0.022\%$ & $0.007\%$ & $0.007\%$ & $\bf 0.003\%$ & $0.009\%$ & $0.017\%$ & $0.035\%$ \\
 $0.68$ & $9.774\%$ & $2.607\%$ & $0.611\%$ & $0.079\%$ & $0.049\%$ & $0.027\%$ & $0.024\%$ & $0.008\%$ & $0.009\%$ & $\bf 0.005\%$ & $0.010\%$ & $0.015\%$ & $0.034\%$ \\
 $0.69$ & $9.756\%$ & $2.576\%$ & $0.619\%$ & $0.077\%$ & $0.053\%$ & $0.029\%$ & $0.028\%$ & $0.012\%$ & $0.015\%$ & $\bf 0.010\%$ & $0.014\%$ & $0.014\%$ & $0.035\%$ \\
 $0.70$ & $9.740\%$ & $2.543\%$ & $0.629\%$ & $0.076\%$ & $0.060\%$ & $0.035\%$ & $0.036\%$ & $0.021\%$ & $ 0.025\%$ & $\bf 0.020\%$ & $  0.025\%$ & $  0.022\%$ & $  0.044\%$ \\
 $0.71$ & $9.727\%$ & $2.509\%$ & $0.642\%$ & $0.079\%$ & $0.072\%$ & $0.047\%$ & $0.050\%$ & $\bf 0.038\%$ & $ 0.044\%$ & $  0.042\%$ & $  0.050\%$ & $  0.050\%$ & $  0.076\%$ \\
 $0.72$ & $9.717\%$ & $2.472\%$ & $0.658\%$ & $0.088\%$ & $0.091\%$ & $\bf 0.069\%$ & $0.075\%$ & $\bf 0.069\%$ & $ 0.080\%$ & $  0.088\%$ & $  0.108\%$ & $  0.123\%$ & $  0.172\%$ \\
 $0.73$ & $9.711\%$ & $2.432\%$ & $0.678\%$ & $0.108\%$ & $0.119\%$ & $\bf 0.105\%$ & $0.118\%$ & $ 0.125\%$ & $ 0.153\%$ & $  0.186\%$ & $  0.241\%$ & $  0.310\%$ & $  0.437\%$ \\
 $0.74$ & $9.708\%$ & $2.391\%$ & $0.705\%$ & $\bf 0.140\%$ & $0.163\%$ & $0.163\%$ & $0.193\%$ & $ 0.229\%$ & $ 0.299\%$ & $  0.401\%$ & $  0.560\%$ & $  0.793\%$ & $  1.190\%$ \\
 $0.75$ & $9.709\%$ & $2.347\%$ & $0.739\%$ & $\bf 0.191\%$ & $0.229\%$ & $0.256\%$ & $0.325\%$ & $ 0.427\%$ & $ 0.603\%$ & $  0.887\%$ & $  1.348\%$ & $  2.088\%$ & $  3.411\%$ \\
 $0.76$ & $9.714\%$ & $2.302\%$ & $0.783\%$ & $\bf 0.264\%$ & $0.330\%$ & $0.409\%$ & $0.561\%$ & $ 0.809\%$ & $ 1.248\%$ & $  2.030\%$ & $  3.374\%$ & $  5.652\%$ & $ 10.214\%$ \\
 $0.77$ & $9.724\%$ & $2.255\%$ & $0.842\%$ & $\bf 0.369\%$ & $0.486\%$ & $0.665\%$ & $0.992\%$ & $ 1.569\%$ & $ 2.665\%$ & $  4.857\%$ & $  8.807\%$ & $ 15.547\%$ & $ 31.198\%$ \\
 $0.78$ & $9.738\%$ & $2.208\%$ & $0.920\%$ & $\bf 0.518\%$ & $0.727\%$ & $1.100\%$ & $1.801\%$ & $ 3.122\%$ & $ 5.886\%$ & $ 12.405\%$ & $ 23.864\%$ & $ 42.057\%$ & $ 88.622\%$ \\
 $0.79$ & $9.757\%$ & $2.164\%$ & $1.027\%$ & $\bf 0.733\%$ & $1.110\%$ & $1.864\%$ & $3.366\%$ & $ 6.373\%$ & $13.443\%$ & $ 36.509\%$ & $ 62.936\%$ & $106.938\%$ & $218.171\%$ \\
 $0.80$ & $9.782\%$ & $2.125\%$ & $1.175\%$ & $\bf 1.046\%$ & $1.730\%$ & $3.250\%$ & $6.505\%$ & $13.306\%$ & $31.383\%$ & $118.467\%$ & $145.999\%$ & $251.290\%$ & $493.722\%$ \\
\hline
\end{tabular}
\caption{Relative error of WKB formula of each order for the massive scalar field ($D=4$, $\ell=1$, $n=0$). The minimal relative error is given in bold. For small mass the minimum error is provided by the WKB formula of 10th order. As asymptotical value approaches the value of peak the minimum error is provided by lower order WKB formula. For larger mass the WKB formula does not work since is does not take into account the additional turning points.}\label{tabl:massivel1}
\end{table*}

For $D=4$ and $\ell=1$ (see table~\ref{tabl:massivel1}) we observe that for small mass the best approximation is given by the 10th-order WKB formula. However, as the squared mass approaches the value of the potential in its maximum, the error of the WKB formula increases, and the best approximation is given by the lower-order formula.

\begin{table*}
\begin{tabular}{|l|r|r|r|r|r|r|r|r|r|r|r|r|}
\hline
 $\mu$ & 2 order & 3 order & 4 order & 5 order & 6 order & 7 order & 8 order & 9 order & 10 order & 11 order & 12 order & 13 order \\
\hline
 $1.20$ & $0.4796\%$ & $0.1018\%$ & $0.0130\%$ & $0.0124\%$ & $0.0075\%$ & $0.0062\%$ & $0.0048\%$ & $0.0041\%$ & $0.0036\%$ & $0.0033\%$ & $0.0031\%$ & $\bf 0.0030\%$ \\
 $1.21$ & $0.4705\%$ & $0.1061\%$ & $0.0167\%$ & $0.0155\%$ & $0.0103\%$ & $0.0088\%$ & $0.0073\%$ & $0.0066\%$ & $0.0061\%$ & $0.0060\%$ & $\bf 0.0060\%$ & $ 0.0061\%$ \\
 $1.22$ & $0.4609\%$ & $0.1111\%$ & $0.0215\%$ & $0.0197\%$ & $0.0142\%$ & $0.0126\%$ & $0.0111\%$ & $0.0107\%$ & $\bf 0.0106\%$ & $0.0109\%$ & $0.0116\%$ & $ 0.0127\%$ \\
 $1.23$ & $0.4508\%$ & $0.1172\%$ & $0.0276\%$ & $0.0252\%$ & $0.0197\%$ & $0.0183\%$ & $\bf 0.0173\%$ & $0.0176\%$ & $0.0185\%$ & $0.0203\%$ & $0.0230\%$ & $ 0.0269\%$ \\
 $1.24$ & $0.4402\%$ & $0.1246\%$ & $0.0353\%$ & $0.0326\%$ & $0.0276\%$ & $\bf 0.0268\%$ & $0.0271\%$ & $0.0293\%$ & $0.0330\%$ & $0.0386\%$ & $0.0468\%$ & $ 0.0584\%$ \\
 $1.25$ & $0.4291\%$ & $0.1335\%$ & $0.0451\%$ & $0.0426\%$ & $\bf 0.0388\%$ & $0.0399\%$ & $0.0432\%$ & $0.0497\%$ & $0.0598\%$ & $0.0749\%$ & $0.0972\%$ & $ 0.1303\%$ \\
 $1.26$ & $0.4177\%$ & $0.1445\%$ & $0.0577\%$ & $0.0562\%$ & $\bf 0.0550\%$ & $0.0601\%$ & $0.0696\%$ & $0.0857\%$ & $0.1106\%$ & $0.1488\%$ & $0.2073\%$ & $ 0.2989\%$ \\
 $1.27$ & $0.4060\%$ & $0.1579\%$ & $\bf 0.0739\%$ & $0.0750\%$ & $0.0787\%$ & $0.0918\%$ & $0.1141\%$ & $0.1507\%$ & $0.2093\%$ & $0.3030\%$ & $0.4544\%$ & $ 0.7075\%$ \\
 $1.28$ & $0.3943\%$ & $0.1744\%$ & $\bf 0.0949\%$ & $0.1010\%$ & $0.1141\%$ & $0.1425\%$ & $0.1903\%$ & $0.2707\%$ & $0.4060\%$ & $0.6345\%$ & $1.0260\%$ & $ 1.7317\%$ \\
 $1.29$ & $0.3831\%$ & $0.1949\%$ & $\bf 0.1223\%$ & $0.1376\%$ & $0.1674\%$ & $0.2247\%$ & $0.3238\%$ & $0.4976\%$ & $0.8095\%$ & $1.3708\%$ & $2.3883\%$ & $ 4.3904\%$ \\
 $1.30$ & $0.3730\%$ & $0.2205\%$ & $\bf 0.1586\%$ & $0.1896\%$ & $0.2492\%$ & $0.3612\%$ & $0.5627\%$ & $0.9383\%$ & $1.6667\%$ & $3.0670\%$ & $5.7214\%$ & $11.5064\%$ \\
\hline
\end{tabular}
\caption{Relative error of WKB formula of each order for the massive scalar field ($D=4$, $\ell=2$, $n=0$). The minimal relative error is given in bold. For small mass the minimum error is provided by the WKB formula of the highest order. As asymptotical value approaches the value of peak the minimum error is provided by lower order WKB formula. For larger mass the WKB formula does not work since is does not take into account the additional turning points.}\label{tabl:massivel2}
\end{table*}

For larger values of the multipole number (see table~\ref{tabl:massivel2} for $\ell=2$) higher orders of the WKB formula can be used to obtain quasinormal modes even for large values of $\mu$. When the asymptotical value of the effective potential increases and approaches the height of the potential barrier, it seems that the best approximation is provided by the WKB formula of 4th order, leading to the relative error almost twice smaller than the 3rd-order formula \cite{Iyer:1986np}.

\begin{table*}
\begin{tabular}{|l|r|r|r|r|r|r|r|r|r|r|}
\hline
 $\mu$ & 4 order & 5 order & 6 order & 7 order & 8 order & 9 order & 10 order & 11 order & 12 order & 13 order \\
\hline
 $0.65$ & $0.06396\%$ & $0.04155\%$ & $0.00901\%$ & $ 0.00623\%$ & $ 0.00168\%$ & $ 0.00147\%$ & $0.00275\%$ & $ 0.00061\%$ & $\bf 0.00034\%$ & $ 0.00038\%$ \\
 $0.66$ & $0.06399\%$ & $0.04625\%$ & $0.00688\%$ & $ 0.00913\%$ & $ 0.00242\%$ & $ 0.00166\%$ & $0.00282\%$ & $ 0.00150\%$ & $\bf 0.00019\%$ & $\bf 0.00025\%$ \\
 $0.67$ & $0.06475\%$ & $0.04874\%$ & $0.00686\%$ & $ 0.01180\%$ & $ 0.00344\%$ & $ 0.00218\%$ & $0.00232\%$ & $ 0.00173\%$ & $\bf 0.00008\%$ & $\bf 0.00024\%$ \\
 $0.68$ & $0.06696\%$ & $0.04721\%$ & $0.01114\%$ & $ 0.01540\%$ & $ 0.00264\%$ & $ 0.00318\%$ & $\bf 0.00218\%$ & $ 0.00179\%$ & $ 0.00235\%$ & $\bf 0.00021\%$ \\
 $0.69$ & $0.07173\%$ & $0.07577\%$ & $0.01989\%$ & $ 0.02014\%$ & $ 0.00354\%$ & $ 0.00480\%$ & $\bf 0.00179\%$ & $ 0.00169\%$ & $ 0.00380\%$ & $\bf 0.00021\%$ \\
 $0.70$ & $0.08050\%$ & $0.08385\%$ & $0.03273\%$ & $ 0.03118\%$ & $ 0.00467\%$ & $ 0.00089\%$ & $\bf 0.00153\%$ & $ 0.00153\%$ & $ 0.00067\%$ & $\bf 0.00033\%$ \\
 $0.71$ & $0.09492\%$ & $0.09396\%$ & $0.04798\%$ & $ 0.03542\%$ & $ 0.00608\%$ & $ 0.00149\%$ & $0.00154\%$ & $ 0.00141\%$ & $ 0.00099\%$ & $\bf 0.00058\%$ \\
 $0.72$ & $0.11675\%$ & $0.10717\%$ & $0.06116\%$ & $ 0.03314\%$ & $ 0.00773\%$ & $ 0.00263\%$ & $0.00161\%$ & $ 0.00138\%$ & $\bf 0.00252\%$ & $\bf 0.00134\%$ \\
 $0.73$ & $0.14793\%$ & $0.13153\%$ & $0.06864\%$ & $ 0.03471\%$ & $ 0.00940\%$ & $ 0.00462\%$ & $0.00195\%$ & $ 0.00168\%$ & $\bf 0.00359\%$ & $\bf 0.00185\%$ \\
 $0.74$ & $0.19088\%$ & $0.15701\%$ & $0.07125\%$ & $ 0.03628\%$ & $ 0.01070\%$ & $ 0.00806\%$ & $0.00302\%$ & $ 0.00278\%$ & $\bf 0.00216\%$ & $\bf 0.00211\%$ \\
 $0.75$ & $0.24872\%$ & $0.19013\%$ & $0.07152\%$ & $ 0.04117\%$ & $ 0.01179\%$ & $ 0.01408\%$ & $0.00269\%$ & $ 0.00501\%$ & $\bf 0.00095\%$ & $ 0.00204\%$ \\
 $0.76$ & $0.27576\%$ & $0.23657\%$ & $0.07130\%$ & $ 0.04198\%$ & $ 0.01531\%$ & $ 0.02193\%$ & $0.00436\%$ & $ 0.00236\%$ & $\bf 0.00112\%$ & $\bf 0.00190\%$ \\
 $0.77$ & $0.36274\%$ & $0.25671\%$ & $0.07213\%$ & $ 0.04465\%$ & $ 0.02565\%$ & $ 0.03423\%$ & $0.00691\%$ & $ 0.00412\%$ & $\bf 0.00137\%$ & $\bf 0.00289\%$ \\
 $0.78$ & $0.47816\%$ & $0.34409\%$ & $0.07638\%$ & $ 0.05096\%$ & $ 0.04419\%$ & $ 0.04838\%$ & $0.01047\%$ & $ 0.00783\%$ & $\bf 0.00315\%$ & $ 0.00653\%$ \\
 $0.79$ & $0.62909\%$ & $0.21752\%$ & $0.08843\%$ & $ 0.11649\%$ & $ 0.05517\%$ & $ 0.04116\%$ & $0.01438\%$ & $ 0.01944\%$ & $\bf 0.00324\%$ & $ 0.01109\%$ \\
 $0.80$ & $0.82357\%$ & $0.20480\%$ & $0.11501\%$ & $ 0.14408\%$ & $ 0.06788\%$ & $ 0.05035\%$ & $0.01848\%$ & $ 0.02618\%$ & $ 0.00666\%$ & $\bf 0.00522\%$ \\
 $0.81$ & $1.06986\%$ & $0.16761\%$ & $0.16460\%$ & $ 0.20743\%$ & $ 0.07275\%$ & $ 0.05513\%$ & $0.03057\%$ & $ 0.04632\%$ & $\bf 0.01214\%$ & $\bf 0.01686\%$ \\
 $0.82$ & $1.37667\%$ & $0.11644\%$ & $0.24827\%$ & $ 0.27202\%$ & $ 0.07815\%$ & $ 0.06577\%$ & $0.05506\%$ & $ 0.07202\%$ & $\bf 0.01889\%$ & $\bf 0.02976\%$ \\
 $0.83$ & $1.75839\%$ & $0.08984\%$ & $0.34336\%$ & $ 0.30013\%$ & $ 0.09740\%$ & $\bf 0.14587\%$ & $0.07643\%$ & $ 0.06695\%$ & $\bf 0.03021\%$ & $ 0.05068\%$ \\
 $0.84$ & $2.25345\%$ & $0.16686\%$ & $\bf 0.53373\%$ & $ 0.26533\%$ & $ 0.15147\%$ & $ 0.22960\%$ & $0.08649\%$ & $ 0.08892\%$ & $\bf 0.05902\%$ & $ 0.09246\%$ \\
 $0.85$ & $1.27000\%$ & $0.33693\%$ & $0.83271\%$ & $ 0.24505\%$ & $ 0.25331\%$ & $ 0.33193\%$ & $0.10499\%$ & $\bf 0.14175\%$ & $\bf 0.08616\%$ & $ 0.09516\%$ \\
 $0.86$ & $1.51651\%$ & $0.61480\%$ & $1.27359\%$ & $ 0.18356\%$ & $ 0.44538\%$ & $ 0.39136\%$ & $0.17105\%$ & $ 0.23958\%$ & $\bf 0.10088\%$ & $ 0.12708\%$ \\
 $0.87$ & $1.85416\%$ & $1.07305\%$ & $1.88320\%$ & $ 0.21607\%$ & $ 0.76016\%$ & $ 0.24206\%$ & $0.33931\%$ & $ 0.36743\%$ & $\bf 0.15757\%$ & $\bf 0.22342\%$ \\
 $0.88$ & $2.29047\%$ & $1.86311\%$ & $1.29375\%$ & $ 0.50983\%$ & $ 1.15363\%$ & $ 0.17200\%$ & $0.68637\%$ & $ 0.43566\%$ & $\bf 0.34491\%$ & $\bf 0.37365\%$ \\
 $0.89$ & $2.83729\%$ & $3.29516\%$ & $1.73169\%$ & $ 1.17081\%$ & $ 1.71926\%$ & $\bf 0.42673\%$ & $1.17958\%$ & $ 0.80329\%$ & $\bf 0.78287\%$ & $ 0.69633\%$ \\
 $0.90$ & $3.49431\%$ & $6.37234\%$ & $2.41981\%$ & $ 2.62674\%$ & $ 3.23255\%$ & $\bf 1.26828\%$ & $2.15320\%$ & $ 3.86396\%$ & $ 1.42861\%$ & $ 2.71284\%$ \\
 $0.91$ & $4.18599\%$ & $6.71647\%$ & $\bf 3.38210\%$ & $ 5.90473\%$ & $10.06010\%$ & $\bf 3.61640\%$ & $7.04142\%$ & $18.89640\%$ & $ 3.94263\%$ & $12.91890\%$ \\
 $0.92$ & $4.66908\%$ & $8.25373\%$ & $\bf 4.41432\%$ & $13.46270\%$ & $47.20960\%$ & $10.04650\%$ & $6.60398\%$ & $13.45280\%$ & $12.23440\%$ & $58.38840\%$ \\
\hline
\end{tabular}
\caption{Relative error of the frequency calculated by averaging results obtained by Padé approximants of each order, from 4th to 13th, for the massive scalar field for ($D=4$, $\ell=1$, $n=0$). The minimal error and the minimal estimation by standard deviation formula are given in bold. We see that the error estimation works well for choosing the best order. Unlike for the ordinary WKB formula, we see that the best approximation is usually provided by higher WKB orders.}
\label{tabl:massivel1Pade}
\end{table*}

On table~\ref{tabl:massivel1Pade} one can see that the above describing averaging of Padé approximations allow one not only to considerably improve accuracy of the quasinormal modes for $\mu^2\lessapprox V_0$, but also calculate the frequencies for $\mu^2 \gtrsim V_0$ when the maximum of the potential still exists (see the potential for $\mu=0.9$ on fig.~\ref{fig:massivescalarpot}). Despite the WKB expansion does not take into account additional turning point the method allows to calculate the dominant frequency with surprisingly good accuracy whenever the potential has a local maximum.  Although the standard deviation formula does not always suggest the WKB order, which gives the best accuracy, it usually allows to choose the WKB order, which provides quite accurate answers. Because our averaging method is based on an ad~hoc choice of the values provided by different Padé approximations our estimation of the error sometimes fails to determine the correct order of error quite frequently. In some cases, the chosen values appear to be much closer to each other than to the accurate frequency, yielding the standard deviation of a couple of orders less than the actual error. That is why a certain amount of caution must be employed when choosing the appropriate WKB order in order to obtain a good approximation.

From table~\ref{tabl:massivel1Pade} one can learn that the actual error changes quite smoothly as a function of the considered parameter, i.~e.~the field mass. At the same time, the estimations of the error with the help of the standard deviation formula may vary a lot for some values of the parameter, leading sometimes to a wrong choice of the best WKB order for the calculation of the quasinormal mode at given value of $\mu$. That is why, when studying how quasinormal modes depend on a parameter, it may be useful to fix the WKB order and use the standard deviation formula to control growth of the error as the parameter changes. The significant growth of the error estimation usually indicates that the considered WKB order might not provide a good approximation.

\begin{table}
\begin{tabular}{|c|c|c|c|c|}
\hline
 $\mu$ & $\omega$ & $\omega_{13}$ & $S$ & $|\omega-\omega_{13}|$ \\
\hline
 $0.00$ & $0.2209-0.2098\imo$ & $0.2213-0.2099\imo$ & $0.000835$ & $0.000410$ \\
 $0.05$ & $0.2212-0.2088\imo$ & $0.2216-0.2088\imo$ & $0.000842$ & $0.000405$ \\
 $0.10$ & $0.2220-0.2057\imo$ & $0.2224-0.2057\imo$ & $0.000873$ & $0.000406$ \\
 $0.15$ & $0.2232-0.2006\imo$ & $0.2239-0.2006\imo$ & $0.001073$ & $0.000662$ \\
 $0.20$ & $0.2247-0.1937\imo$ & $0.2251-0.1939\imo$ & $0.001546$ & $0.000454$ \\
 $0.25$ & $0.2264-0.1850\imo$ & $0.2274-0.1857\imo$ & $0.001578$ & $0.001287$ \\
 $0.30$ & $0.2282-0.1748\imo$ & $0.2341-0.1717\imo$ & $0.004243$ & $0.006653$ \\
 $0.35$ & $0.2302-0.1634\imo$ & $0.2451-0.1561\imo$ & $0.018104$ & $0.016651$ \\
 $0.40$ & $0.2324-0.1507\imo$ & $0.2321-0.1330\imo$ & $0.012704$ & $0.017702$ \\
 $0.45$ & $0.2350-0.1369\imo$ & $0.2536-0.1354\imo$ & $0.051395$ & $0.018657$ \\
\hline
\end{tabular}
\caption{Fundamental frequency for the massive scalar field for ($D=4$, $\ell=0$, $n=0$) compared with the frequency calculated by averaging results obtained by Padé approximants of 13th order. The standard deviation formula provides good estimates for the absolute error for all $\mu$.}
\label{tabl:massivel0}
\end{table}

In this way we used averaging of the Padé approximations of 13th order to calculate the fundamental frequency for the massive scalar field as a function of the field mass (see table~\ref{tabl:massivel0}). Rapid growth of the error estimation indicates that for $\mu\gtrsim0.3$ the WKB approach becomes less accurate (in this case due to an additional turning point). In this particular case the standard deviation formula correctly estimates the absolute error for all $\mu$, however, this does not happen for all WKB orders. If one considered the best order suggested by the standard deviation formula for each value of $\mu$ separately, the error estimations would be incorrect and the results would be less accurate. We believe that looking at the error estimation as a function of the parameter for each order may give a hint for the appropriate choice of the WKB order in the particular parametric range.

\subsection{Scattering problem}

Our tests of the higher-order WKB formula for the scattering of a massless scalar field show that the ordinary WKB formula (\ref{WKBformula-spherical}) provides a good approximation for the transmission coefficient for any $\ell>0$.

\begin{figure*}
\resizebox{\linewidth}{!}{\includegraphics*{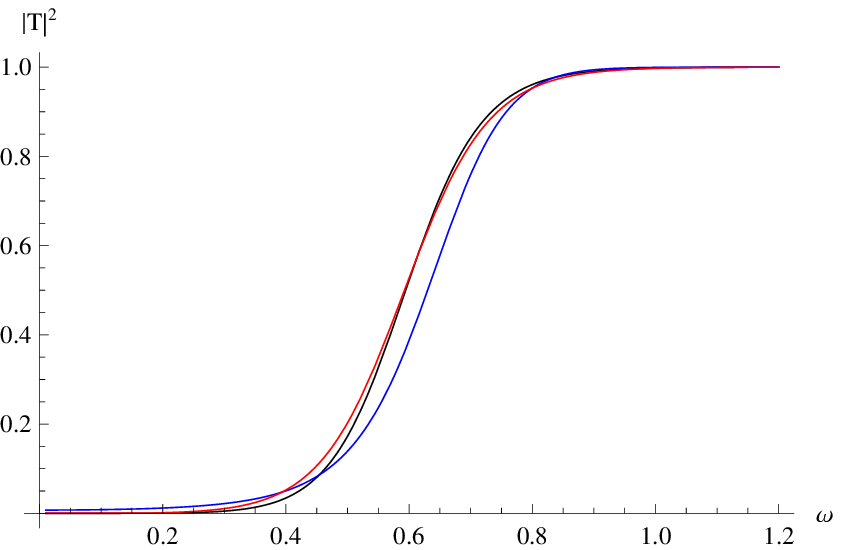}\includegraphics*{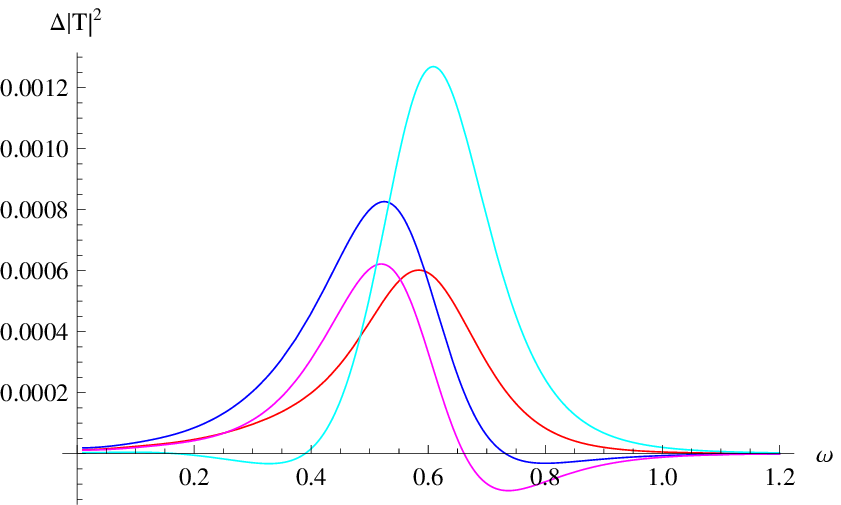}}
\caption{Left panel: Transmission coefficients for the massive scalar field for $D=4$, $\ell=1$ (black), eikonal approximation (blue), second-order WKB formula (red). Right panel: errors for the WKB formula of 6th order, $|T_6|^2-|T|^2$ (blue), and 7th order $|T_7|^2-|T|^2$ (red), and estimations given by $|T_7|^2-|T_5|^2$ (cyan) and $|T_6|^2-|T_8|^2$ (magenta).}\label{fig:Transmission}
\end{figure*}

On Fig.~\ref{fig:Transmission} we show that the 6th and 7th WKB orders allow one to calculate the transmission coefficient already for $\ell=1$ with the maximal error of about $0.1\%$. Comparison of the results obtained using different WKB orders gives a good estimation of the error again. The accuracy for higher values of $\ell$ is even better, so that it is possible to get a good approximation using a low-order WKB formula.

For $\ell=0$ the WKB approach described in Sec.~\ref{sec:WKB} encounters a few problems. First, the eikonal formula does not give a good estimation for the transmission coefficient, except for large $\omega$, when $|T|^2\approx1$. This complicates the appropriate choice of $\K$ among the solutions to the equation (\ref{WKBformula}), when the WKB order is large. Second, increasing of the WKB order does not improve the accuracy significantly. The reason is that the increasing of the WKB order leads to growing of the error for small $\omega$, while improving accuracy for intermediate values of $\omega$. Thus, for $D=4$, $\ell=0$ the overall error for the transmission coefficient remains of the order of several percents.

One could expect that usage of the Padé approximants in a similar way as for the quasinormal modes could improve the accuracy of WKB grey-body factors as well. In order to check this assumption, we have studied solutions to the equation (\ref{omegaPade}) with respect to $\K$. It turns out that, for the equation with a given choice of the Padé approximant, it is not always possible to find a solution $\K$ which is the nearest to the eikonal one. In some cases the corresponding equation (\ref{omegaPade}) does not even have any solution with vanishing real part. However, by comparing the accurate value of $K$ with all purely imaginary solutions to all possible equations with Padé approximations of a given order, we learned that the most accurate of such solutions has smaller error than the solution obtained with the help of the ordinary WKB formula. This observation suggests that employing Padé approximants may indeed improve the accuracy of the transmission coefficient at high WKB orders. Yet, due to computational difficulties, we were limited by relatively small (4th and 5th) WKB orders, for which improvement of the accuracy is not very strong.

Unfortunately, when the coefficients in (\ref{WKBPade}) depend on $\K$, deriving the Padé approximants is a very time and memory consuming procedure at large orders. The \emph{Mathematica®} built-in algorithm allows one to perform such operations for a reasonable time on a personal computer when the order is not higher than six \cite{Mathematica-code}. That is why even assuming that one finds out how to select the best value for $\K$ among all the possible solutions, we doubt that such a method based on the equation (\ref{omegaPade}) would be practically useful.

\section{Conclusions}\label{sec:summary}

We have reviewed the state-of-art of the WKB approach applied to the quasinormal modes and grey-body factors of black holes. A special attention has been given to the improvement of the WKB approach using Padé approximants suggested in \cite{Matyjasek:2017psv}. We have proposed an ad~hoc averaging method for the Padé approximations which enables us to calculate quasinormal modes with much better accuracy comparing to the ordinary WKB formula. Using this method we are able to calculate not only dominant, but also all practically relevant overtones of the gravitational perturbation spectra \cite{Giesler:2019uxc} with a reasonable accuracy. In addition, we can estimate an error of the approximation within a given WKB order. In the attachment to this paper we share with readers an automatic \emph{Mathematica®} code \cite{Mathematica-code}, which can be used for calculation of the quasinormal modes and grey-body factors of black holes using the WKB method.

The WKB approach leaves a number of essential open questions:
\begin{itemize}
\item We have suggested a procedure for averaging of the Padé approximations of a given order, which allows one to improve the accuracy of the WKB approximation considerably. However, this method does not always give a reliable result and neither guarantee a good estimation for the error. We believe that development of a mathematically strict and universal method of calculation of quasinormal modes based on the Padé approximation would give us a very powerful tool to study various spectral problems for black holes.
\item The method of Padé approximations has been formulated only for the particular type of the effective potential, $U(x,\omega)=V(x)-\omega^2$. Proper generalization of this approach for the arbitrary form of $U(x,\omega)$ could improve accuracy of the WKB method for axisymmetric black holes, and, perhaps, give some insights on general applicability of the WKB approach.
\item While the ordinary WKB formula usually gives the best approximation at some finite order, the method of Padé approximations considerably improves the accuracy of the WKB method of higher orders. This motivates calculation of explicit expressions for the WKB corrections of orders higher than 13.
\item The scattering problem for the Schwarzschild black hole can be solved using the ordinary WKB formula for $\ell>0$. In this way all physically relevant grey-body factors can be computed within the WKB approach. However, for $\ell=0$ as well as for more complicated potentials the WKB formula usually does not provide an adequate accuracy. At the same time, it turns out that the straightforward using of the method of Padé approximations is a time consuming method. Besides, it is not clear how to select the most accurate solutions among the obtained values. Development of a robust algorithm for this case would make the WKB approach a universal tool for studying of black-hole perturbations.
\item Another interesting question in this context is a generalization of the formulas (\ref{reflection}) and (\ref{transmission}) for any complex-valued potential, which would allow one to study scattering near the axisymmetric black holes with the help of the WKB method.
\end{itemize}

\begin{acknowledgments}
The authors would like to acknowledge J.~Matyjasek and M.~Opala for the sharing their Mathematica® notebook with higher order WKB corrections \cite{Matyjasek:2017psv}. We also thank J.~Matyjasek, M.~Opala, and B.~Mashhoon for useful discussions.

The authors acknowledge the support of the grant 19-03950S of Czech Science Foundation (GAČR). This publication has been prepared with the support of the “RUDN University Program 5-100”. AFZ acknowledges the SU grant SGS/12/2019.
\end{acknowledgments}


\begin{thebibliography}{99}
\bibitem{1985ApJ...291L..33S} B.~F.~Schutz and C.~M.~Will, Astrophys.\ J.\ Lett.\ {\bf 291:} L33-36 (1985).

\bibitem{Iyer:1986np}
  S.~Iyer and C.~M.~Will,
  Phys.\ Rev.\ D {\bf 35}, 3621 (1987).

\bibitem{Konoplya:2003ii}
  R.~A.~Konoplya,
  Phys.\ Rev.\ D {\bf 68}, 024018 (2003)
  [gr-qc/0303052];
  J.\ Phys.\ Stud.\  {\bf 8}, 93 (2004).

\bibitem{Matyjasek:2017psv}
  J.~Matyjasek and M.~Opala,
  Phys.\ Rev.\ D {\bf 96}, no. 2, 024011 (2017)
  [arXiv:1704.00361 [gr-qc]].

\bibitem{Mathematica-code} The \emph{Mathematica®} package with the WKB formula of 13th order and Padé approximations ready for calculation of the quasinormal modes and grey-body factors, as well as examples of such calculations for the Schwarzschild black hole are publicly available to download from \url{https://goo.gl/nykYGL}.

\bibitem{TheLIGOScientific:2016src}
  B.~P.~Abbott {\it et al.} [LIGO Scientific and Virgo Collaborations],
  Phys.\ Rev.\ Lett.\  {\bf 116}, no. 6, 061102 (2016)
  [arXiv:1602.03837 [gr-qc]];
  Phys.\ Rev.\ Lett.\  {\bf 116}, no. 22, 221101 (2016)
  [arXiv:1602.03841 [gr-qc]];
  Phys.\ Rev.\ Lett.\  {\bf 116}, no. 24, 241103 (2016)
  [arXiv:1606.04855 [gr-qc]].

\bibitem{Konoplya:2011qq}
  R.~A.~Konoplya and A.~Zhidenko,
  Rev.\ Mod.\ Phys.\  {\bf 83}, 793 (2011)
  [arXiv:1102.4014 [gr-qc]].

\bibitem{Mashhoon}
B.~Mashhoon, Proc.\ Third Marcel Grossmann Meeting on General Relativity ed Hu Ning (Amsterdam: North-Holland) pp 599–608 (1983);
H.-J.~Blome and B.~Mashhoon, Phys.\ Lett.\ A110, 231 (1984);
H.~Liu and B.~Mashhoon, Class.\ Quant.\ Grav.\  {\bf 13} 233–251 (1996).

\bibitem{Poschl-Teller} G.~Pöschl and E.~Teller, Z.\ Physik\ {\bf 83}, 143 (1933).

\bibitem{Konoplya:2001ji}
  R.~A.~Konoplya,
  Gen.\ Rel.\ Grav.\  {\bf 34}, 329 (2002)
  [gr-qc/0109096].

\bibitem{Froeman:1992gp}
  N.~Froeman, P.~O.~Froeman, N.~Andersson and A.~Hoekback,
  Phys.\ Rev.\ D {\bf 45}, 2609 (1992).

\bibitem{Andersson:1992scr}
  N.~Andersson and S.~Linnæus,
  Phys.\ Rev.\ D {\bf 46}, no. 10, 4179 (1992).

\bibitem{Galtsov:1991nwq}
  D.~V.~Gal'tsov and A.~A.~Matiukhin,
  Class.\ Quant.\ Grav.\  {\bf 9}, 2039 (1992).

\bibitem{Simone:1991wn}
  L.~E.~Simone and C.~M.~Will,
  Class.\ Quant.\ Grav.\  {\bf 9}, 963 (1992).

\bibitem{Konoplya:2010kv}
  R.~A.~Konoplya and A.~Zhidenko,
  Phys.\ Rev.\ D {\bf 81}, 124036 (2010)
  [arXiv:1004.1284 [hep-th]];
  Phys.\ Rev.\ D {\bf 82}, 084003 (2010)
  [arXiv:1004.3772 [hep-th]].


\bibitem{Fernando:2016ftj}
  S.~Fernando,
  Gen.\ Rel.\ Grav.\  {\bf 48}, no. 3, 24 (2016)
  [arXiv:1601.06407 [gr-qc]].

\bibitem{Konoplya:2016pmh}
  R.~Konoplya and A.~Zhidenko,
  Phys.\ Lett.\ B {\bf 756}, 350 (2016)
  [arXiv:1602.04738 [gr-qc]].
  
\bibitem{Molina:2016tkr}
  C.~Molina, A.~B.~Pavan and T.~E.~Medina Torrejón,
  Phys.\ Rev.\ D {\bf 93}, no. 12, 124068 (2016)
  [arXiv:1604.02461 [gr-qc]].
  
\bibitem{Cuyubamba:2016cug}
  M.~A.~Cuyubamba, R.~A.~Konoplya and A.~Zhidenko,
  Phys.\ Rev.\ D {\bf 93}, no. 10, 104053 (2016)
  [arXiv:1604.03604 [gr-qc]].
  
\bibitem{Abbasvandi:2016yos}
  N.~Abbasvandi, M.~J.~Soleimani, W.~A.~T.~Wan Abdullah and S.~Radiman,
  JCAP {\bf 1703}, no. 03, 030 (2017)
  [arXiv:1604.06868 [hep-th]].
  
\bibitem{Toshmatov:2016bsb}
  B.~Toshmatov, Z.~Stuchlík, J.~Schee and B.~Ahmedov,
  Phys.\ Rev.\ D {\bf 93}, no. 12, 124017 (2016)
  [arXiv:1605.02058 [gr-qc]].

\bibitem{Chen:2016qii}
  C.-H.~Chen, H.~T.~Cho, A.~S.~Cornell and G.~Harmsen,
  Phys.\ Rev.\ D {\bf 94}, no. 4, 044052 (2016)
  [arXiv:1605.05263 [gr-qc]].
  
\bibitem{Konoplya:2016hmd}
  R.~A.~Konoplya and A.~Zhidenko,
  JCAP {\bf 1612}, no. 12, 043 (2016)
  [arXiv:1606.00517 [gr-qc]].

\bibitem{Fernando:2017qrd}
  S.~Fernando and A.~Manning,
  Int.\ J.\ Mod.\ Phys.\ D {\bf 26}, no. 09, 1750100 (2017)
  [arXiv:1701.01983 [gr-qc]].

\bibitem{Wahlang:2017zvk}
  W.~Wahlang, P.~A.~Jeena and S.~Chakrabarti,
  Int.\ J.\ Mod.\ Phys.\ D {\bf 26}, no. 14, 1750160 (2017)
  [arXiv:1703.04286 [gr-qc]].

\bibitem{Breton:2017hwe}
  N.~Bretón, T.~Clark and S.~Fernando,
  Int.\ J.\ Mod.\ Phys.\ D {\bf 26}, no. 10, 1750112 (2017)
  [arXiv:1703.10070 [gr-qc]].

\bibitem{Toshmatov:2017bpx}
  B.~Toshmatov, C.~Bambi, B.~Ahmedov, Z.~Stuchlík and J.~Schee,
  Phys.\ Rev.\ D {\bf 96}, 064028 (2017);
  [arXiv:1705.03654 [gr-qc]].

\bibitem{Toshmatov:2017qrq}
  B.~Toshmatov and Z.~Stuchlík,
  Eur.\ Phys.\ J.\ Plus {\bf 132}, no. 7, 324 (2017)
  [arXiv:1707.07419 [gr-qc]].

\bibitem{Ciric:2017rnf}
  M.~D.~Ćirić, N.~Konjik and A.~Samsarov,
  Class.\ Quant.\ Grav.\  {\bf 35}, no. 17, 175005 (2018)
  [arXiv:1708.04066 [hep-th]].
  
\bibitem{Burikham:2017gdm}
  P.~Burikham, S.~Ponglertsakul and L.~Tannukij,
  Phys.\ Rev.\ D {\bf 96}, no. 12, 124001 (2017)
  [arXiv:1709.02716 [gr-qc]].
  
\bibitem{Ovgun:2017dvs}
  A.~Övgün, İ.~Sakallı and J.~Saavedra,
  Chin.\ Phys.\ C {\bf 42}, no. 10, 105102 (2018)
  [arXiv:1708.08331 [physics.gen-ph]].
  
\bibitem{Blazquez-Salcedo:2017bld}
  J.~L.~Blázquez-Salcedo and C.~Knoll,
  Phys.\ Rev.\ D {\bf 97}, no. 4, 044020 (2018)
  [arXiv:1709.07864 [gr-qc]].

\bibitem{Chen:2017kqa}
  C.-H.~Chen, H.~T.~Cho, A.~S.~Cornell, G.~Harmsen and X.~Ngcobo,
  Phys.\ Rev.\ D {\bf 97}, no. 2, 024038 (2018)
  [arXiv:1710.08024 [gr-qc]].
  
\bibitem{Panotopoulos:2017hns}
  G.~Panotopoulos and Á.~Rincón,
  Int.\ J.\ Mod.\ Phys.\ D {\bf 27}, no. 03, 1850034 (2017)
  [arXiv:1711.04146 [hep-th]].
  
\bibitem{MoraisGraca:2017hrf}
  J.~P.~Morais Graça and I.~P.~Lobo,
  Eur.\ Phys.\ J.\ C {\bf 78}, no. 2, 101 (2018)
  [arXiv:1711.08714 [gr-qc]].

\bibitem{Ponglertsakul:2018smo}
  S.~Ponglertsakul, P.~Burikham and L.~Tannukij,
  Eur.\ Phys.\ J.\ C {\bf 78}, no. 7, 584 (2018)
  [arXiv:1803.09078 [gr-qc]].

\bibitem{Wu:2018xza}
  C.~Wu,
  Eur.\ Phys.\ J.\ C {\bf 78}, no. 4, 283 (2018).
  
\bibitem{Toshmatov:2018tyo}
  B.~Toshmatov, Z.~Stuchlík, J.~Schee and B.~Ahmedov,
  Phys.\ Rev.\ D {\bf 97}, no. 8, 084058 (2018)
  [arXiv:1805.00240 [gr-qc]].
  
\bibitem{Konoplya:2018ala}
  R.~A.~Konoplya,
  Phys.\ Lett.\ B {\bf 784}, 43 (2018)
  [arXiv:1805.04718 [gr-qc]].

\bibitem{Panotopoulos:2018can}
  G.~Panotopoulos,
  Gen.\ Rel.\ Grav.\  {\bf 50}, no. 6, 59 (2018)
  [arXiv:1805.04743 [hep-th]].
  
\bibitem{Blazquez-Salcedo:2018ipc}
  J.~L.~Blázquez-Salcedo, X.~Y.~Chew and J.~Kunz,
  Phys.\ Rev.\ D {\bf 98}, no. 4, 044035 (2018)
  [arXiv:1806.03282 [gr-qc]].
  
\bibitem{Chen:2018mkf}
  C.~Y.~Chen and P.~Chen,
  Phys.\ Rev.\ D {\bf 98}, no. 4, 044042 (2018)
  [arXiv:1806.09500 [gr-qc]].
  
\bibitem{Panotopoulos:2018hua}
  G.~Panotopoulos,
  Mod.\ Phys.\ Lett.\ A {\bf 33}, no. 23, 1850130 (2018)
  [arXiv:1807.03278 [gr-qc]].

\bibitem{Das:2018fzc}
  K.~Das, S.~Pramanik and S.~Ghosh,
  Phys.\ Rev.\ D {\bf 99}, no. 2, 024039 (2019)
  [arXiv:1807.08517 [hep-th]].

\bibitem{Dey:2018cws}
  S.~Dey and S.~Chakrabarti,
  Eur.\ Phys.\ J.\ C {\bf 79}, no. 6, 504 (2019)
  [arXiv:1807.09065 [gr-qc]].

\bibitem{Zinhailo:2018ska}
  A.~F.~Zinhailo,
  Eur.\ Phys.\ J.\ C {\bf 78}, no. 12, 992 (2018)
  [Eur.\ Phys.\ J.\  {\bf 78}, 992 (2018)]
  [arXiv:1809.03913 [gr-qc]].

\bibitem{Ge:2018vjq}
  B.~Ge, J.~Jiang, B.~Wang, H.~Zhang and Z.~Zhong,
  JHEP {\bf 1901}, 123 (2019)
  [arXiv:1810.12128 [gr-qc]].
  
\bibitem{Chakrabarty:2018skk}
  H.~Chakrabarty, A.~A.~Abdujabbarov and C.~Bambi,
  Eur.\ Phys.\ J.\ C {\bf 79}, no. 3, 179 (2019)
  [arXiv:1811.02847 [gr-qc]].
  
\bibitem{Chakrabarti:2018aqm}
  S.~Chakrabarti, S.~Chougule and D.~Maity,
  arXiv:1811.09247 [gr-qc].

\bibitem{Oliveira:2018oha}
  R.~Oliveira, D.~M.~Dantas, V.~Santos and C.~A.~S.~Almeida,
  Class.\ Quant.\ Grav.\  {\bf 36}, no. 10, 105013 (2019)
  [arXiv:1812.01798 [gr-qc]].

\bibitem{Chen:2019iuo}
  C.~Y.~Chen and P.~Chen,
  Phys.\ Rev.\ D {\bf 99}, no. 10, 104003 (2019)
  [arXiv:1902.01678 [gr-qc]].

\bibitem{Ding:2019tvs}
  C.~Ding,
  Nucl.\ Phys.\ B {\bf 938}, 736 (2019)
  [arXiv:1812.07994 [gr-qc]].
    
\bibitem{Volkel:2019ahb}
  S.~H.~Völkel, R.~Konoplya and K.~D.~Kokkotas,
  Phys.\ Rev.\ D {\bf 99}, no. 10, 104025 (2019)
  [arXiv:1902.07611 [gr-qc]].

\bibitem{Konoplya:2019hml}
  R.~A.~Konoplya, A.~F.~Zinhailo and Z.~Stuchlík,
  Phys.\ Rev.\ D {\bf 99}, no. 12, 124042 (2019)
  [arXiv:1903.03483 [gr-qc]].

\bibitem{Ciric:2019uab}
  M.~D.~Ćirić, N.~Konjik and A.~Samsarov,
  arXiv:1904.04053 [hep-th].
  
\bibitem{Konoplya:2019ppy}
  R.~A.~Konoplya and A.~F.~Zinhailo,
  Phys.\ Rev.\ D {\bf 99}, no. 10, 104060 (2019)
  [arXiv:1904.05341 [gr-qc]].

\bibitem{Pramanik:2019qgy}
  S.~Pramanik and K.~Das,
  arXiv:1904.07703 [gr-qc].
  
\bibitem{Yabana}
  K.~Yabana and H.~Horiuchi,
  Prog.\ Theor.\ Phys.\  {\bf 71}, 1275 (1984).

\bibitem{Bennet}
 J.~A.~Bennett,
  Math.\ Proc.\ Cambridge\ Phil.\ Soc. {\bf 80(3)}, 527 (1976).

\bibitem{Konoplya:2002ky}
  R.~A.~Konoplya,
  Phys.\ Rev.\ D {\bf 66}, 084007 (2002)
  [gr-qc/0207028];
  Phys.\ Lett.\ B {\bf 550}, 117 (2002)
  [gr-qc/0210105].

\bibitem{PadeApproximation}
L.~Wuytack (Ed.), Padé Approximation and Its Applications: Proceedings of a Conference held in Antwerp, Belgium, 1979. Lecture Notes in Mathematics. Berlin-Heidelberg-New York, Springer-Verlag 1979.

\bibitem{Starobinsky:1973aij}
  A.~A.~Starobinsky,
  Sov.\ Phys.\ JETP {\bf 37}, no. 1, 28 (1973)
  [Zh.\ Eksp.\ Teor.\ Fiz.\  {\bf 64}, 48 (1973)].

\bibitem{Bekenstein:1973mi}
  J.~D.~Bekenstein,
  Phys.\ Rev.\ D {\bf 7}, 949 (1973).

\bibitem{Brito:2015oca}
  R.~Brito, V.~Cardoso and P.~Pani,
  Lect.\ Notes Phys.\  {\bf 906}, pp.1 (2015)
  [arXiv:1501.06570 [gr-qc]].

\bibitem{Ohashi:2004wr}
  A.~Ohashi and M.~a.~Sakagami,
  Class.\ Quant.\ Grav.\  {\bf 21}, 3973 (2004)
  [gr-qc/0407009].

\bibitem{Konoplya:2004wg}
  R.~A.~Konoplya and A.~Zhidenko,
  Phys.\ Lett.\ B {\bf 609}, 377 (2005)
  [gr-qc/0411059].

\bibitem{Hartnoll:2008kx}
  S.~A.~Hartnoll, C.~P.~Herzog and G.~T.~Horowitz,
  JHEP {\bf 0812}, 015 (2008)
  [arXiv:0810.1563 [hep-th]].

\bibitem{Konoplya:2009hv}
  R.~A.~Konoplya and A.~Zhidenko,
  Phys.\ Lett.\ B {\bf 686}, 199 (2010)
  [arXiv:0909.2138 [hep-th]].

\bibitem{Tangherlini:1963bw}
  F.~R.~Tangherlini,
  Nuovo Cim.\  {\bf 27}, 636 (1963).

\bibitem{Rostworowski:2006bp}
  A.~Rostworowski,
  Acta Phys.\ Polon.\ B {\bf 38}, 81 (2007)
  [gr-qc/0606110].

\bibitem{Konoplya:2017tvu}
  R.~A.~Konoplya and A.~Zhidenko,
  Phys.\ Rev.\ D {\bf 97}, no. 8, 084034 (2018)
  [arXiv:1712.06667 [gr-qc]].

\bibitem{Zhang:2018jgj}
  M.~Zhang, J.~Jiang and Z.~Zhong,
  Phys.\ Lett.\ B {\bf 789}, 13 (2019)
  [arXiv:1811.04183 [gr-qc]].

\bibitem{Zhidenko:2006rs}
  A.~Zhidenko,
  Phys.\ Rev.\ D {\bf 74}, 064017 (2006)
  [gr-qc/0607133].

\bibitem{Giesler:2019uxc}
  M.~Giesler, M.~Isi, M.~Scheel and S.~Teukolsky,
  arXiv:1903.08284 [gr-qc].
\end{thebibliography}
\end{document}